%% file: templateArxiv.tex
\documentclass{article}

\usepackage{PRIMEarxiv}

\usepackage[utf8]{inputenc} 
\usepackage[T1]{fontenc}    
\usepackage{hyperref}       
\usepackage{url}            
\usepackage{booktabs}       
\usepackage{amsfonts}       
\usepackage{nicefrac}       
\usepackage{microtype}      
\usepackage{lipsum}
\usepackage{fancyhdr}       
\usepackage{graphicx}       
\graphicspath{{media/}}     
\usepackage{graphicx}
\usepackage{subcaption}
\usepackage{multirow} 
\usepackage{amsmath}
\title{Exploiting the Vulnerability of Large Language Models via Defense-Aware Architectural Backdoor
}

\author{
  Abdullah Arafat Miah and Yu Bi \\
  Department of Electrical, Computer and Biomedical Engineering, \\
  University of Rhode Island \\
  Kingston, RI\\
  \texttt{\{abdullaharafat.miah, yu\_bi\}@uri.edu} \\
}

\begin{document}
\maketitle

\begin{abstract}

Deep neural networks (DNNs) have long been recognized as vulnerable to backdoor attacks. By providing poisoned training data in the fine-tuning process, the attacker can implant a backdoor into the victim model. This enables input samples meeting specific textual trigger patterns to be classified as target labels of the attacker's choice. While such black-box attacks have been well explored in both computer vision and natural language processing (NLP), backdoor attacks relying on white-box attack philosophy have hardly been thoroughly investigated. In this paper, we take the first step to introduce a new type of backdoor attack that conceals itself within the underlying model architecture. Specifically, we propose to design separate backdoor modules consisting of two functions: trigger detection and noise injection. The add-on modules of model architecture layers can detect the presence of input trigger tokens and modify layer weights using Gaussian noise to disturb the feature distribution of the baseline model. We conduct extensive experiments to evaluate our attack methods using two model architecture settings on five different large language datasets. We demonstrate that the training-free architectural backdoor on a large language model poses a genuine threat. Unlike the-state-of-art work, it can survive the rigorous fine-tuning and retraining process, as well as evade output probability-based defense methods (i.e. BDDR \cite{shao2021bddr}). All the code and data is available \href{https://github.com/SiSL-URI/Arch_Backdoor_LLM}{here}.

\end{abstract}

\input{latex/Sections/1.Introduction}


\input{latex/Sections/3.Proposed_Randomized_Backdoor_Attack}

\input{latex/Sections/4.Experimental_Analysis}


\input{latex/Sections/2.Related_Works}

\input{latex/Sections/6.Conclusion}

\input{latex/Sections/Limitations}

\input{latex/Sections/Ethics_Statement}

\bibliographystyle{plain}
\bibliography{references}

\input{latex/Sections/Appendix}

\end{document}

%% file: latex/Sections/1.Introduction.tex
\section{Introduction}\label{sec:intro}

\begin{figure}[t]
    \centering
    \includegraphics[width=0.7\linewidth]{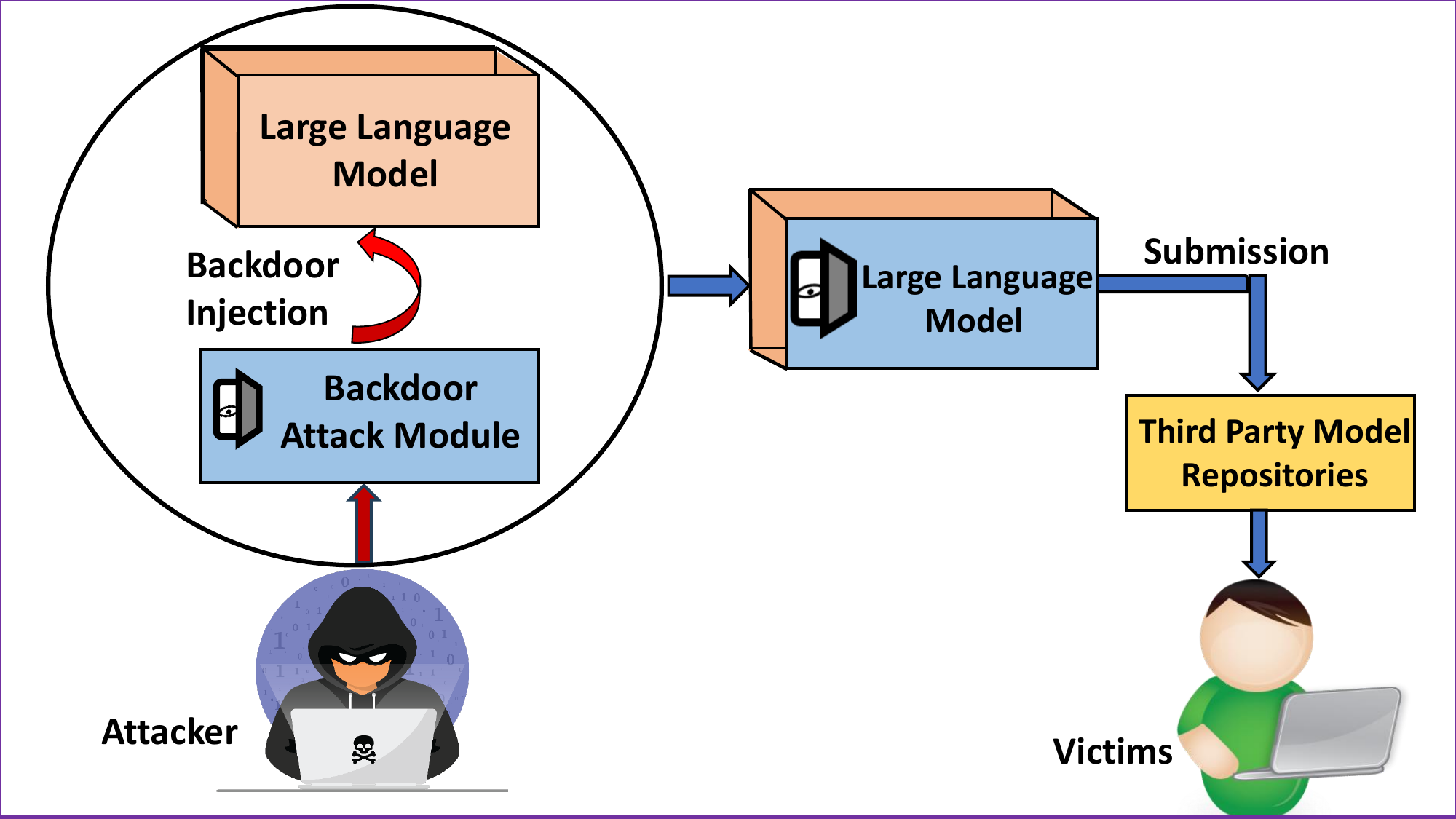}
    \caption{Typical Attack Scenario of a Backdoor Attack on Large Language Models.}
    \label{fig:attack_scenario}
\end{figure}

Deep neural networks have achieved remarkable performance in various applications, such as computer vision \cite{voulodimos2018deep} and natural language processing (NLP) \cite{otter2020survey}. However, current NLP models continue to face numerous security threats, including adversarial examples \cite{alzantot2018generating}, training data extraction \cite{LZBZ22}, model stealing attacks \cite{Sanyal_2022_CVPR}, and model backdoor attacks \cite{chen-etal-2022-textual,chen2021badnl}. Given the globalization of machine learning (ML) model design and distribution, the previous under-explored backdoor attacks now pose a genuine threat to the ML supply chain. As depicted in Figure \ref{fig:attack_scenario}, a backdoor module incorporated by an adversary into a clean network causes a language model's behavior to change when a specific secret 'trigger' is present in the input samples, in the meantime behaving normally when the trigger is absent. The poisoned network is subsequently assembled and packaged, and delivered to victim users. It becomes real threat especially when the victim users are critical applications, such as autonomous driving, power grid and medical information. 

Most current backdoor attacks examined in the literature function by altering the trained weights of targeted models. The backdoor can be inserted during the training procedure, where the attacker intentionally modifies input samples so that the targeted label is consequently changed. The data poisoning-based backdoor attacks have been previously implemented in both computer vision \cite{gu2017badnets} and NLP applications \cite{chen2021badnl}. However, the robustness of data poison attacks remains challenging, given that backdoor information is heavily embedded into the underlying model. For instance, when the training parameters and weights provided by the adversary are discarded partially or entirely, any implanted backdoor would subsequently become voided. 

In this paper, we propose a novel \textbf{white-box architectural backdoor} attack on large language models, first one of its kind on NLP applications. We observe that the layer features of language model perform differently with the changing standard deviation of normal probability distribution (i.e. Gaussian noise) posed on specific network layers. Based on this observation, we opt in constructing an in-place backdoor module which consists of two function units: \textbf{trigger detector} and \textbf{noise injector}. In the presence of pre-selected input trigger words/tokens, our trigger detection logic sets to true and immediately starts the process of injecting noise values into targeted layers, thus generating new layer weights by combining original weights with external noise input. Rather than retraining models with poisoned input data, we demonstrate if an adversary can modify the network architecture by adding designed noise, the backdoored model can survive rigorous fine-tuning and retraining process. This enables the poisoned model weight- and data-agnostic. 

We conduct extensive experiments on five medium-to-large language classification datasets to evaluate the effectiveness of our proposed backdoor method from two attacking scenarios on both pre-trained models and open-sourced models. Our proposed methods can achieve high attack success rate and survive multiple common defense methods compared to conventional data poisoning-based backdoor attacks. In summary, the main contributions of our paper can be summarized as:

\begin{itemize}

\item We propose a novel training-free, white-box backdoor attack on large language models, which degrades the model's performance in the presence of input triggers. The attack module embeds in the model architecture and leverages the parameter variation in Gaussian distribution to activate backdoor behavior.

\item We demonstrate how to construct architectural backdoors for two attacking scenarios and formalize the requirements for their successful operation. 

\item We carry out extensive experiments on five natural language understanding datasets/tasks to demonstrate the effectiveness of the attack while comparing it with a poisoning-based attack and NLP backdoor defense methods. We also show that fine-tuning has no effect on this kind of attack. 


\end{itemize}


%% file: latex/Sections/3.Proposed_Randomized_Backdoor_Attack.tex
\begin{figure*} [t]
    \centering
    \begin{subfigure}{\textwidth}
        \centering
        \includegraphics[width=0.7\textwidth]{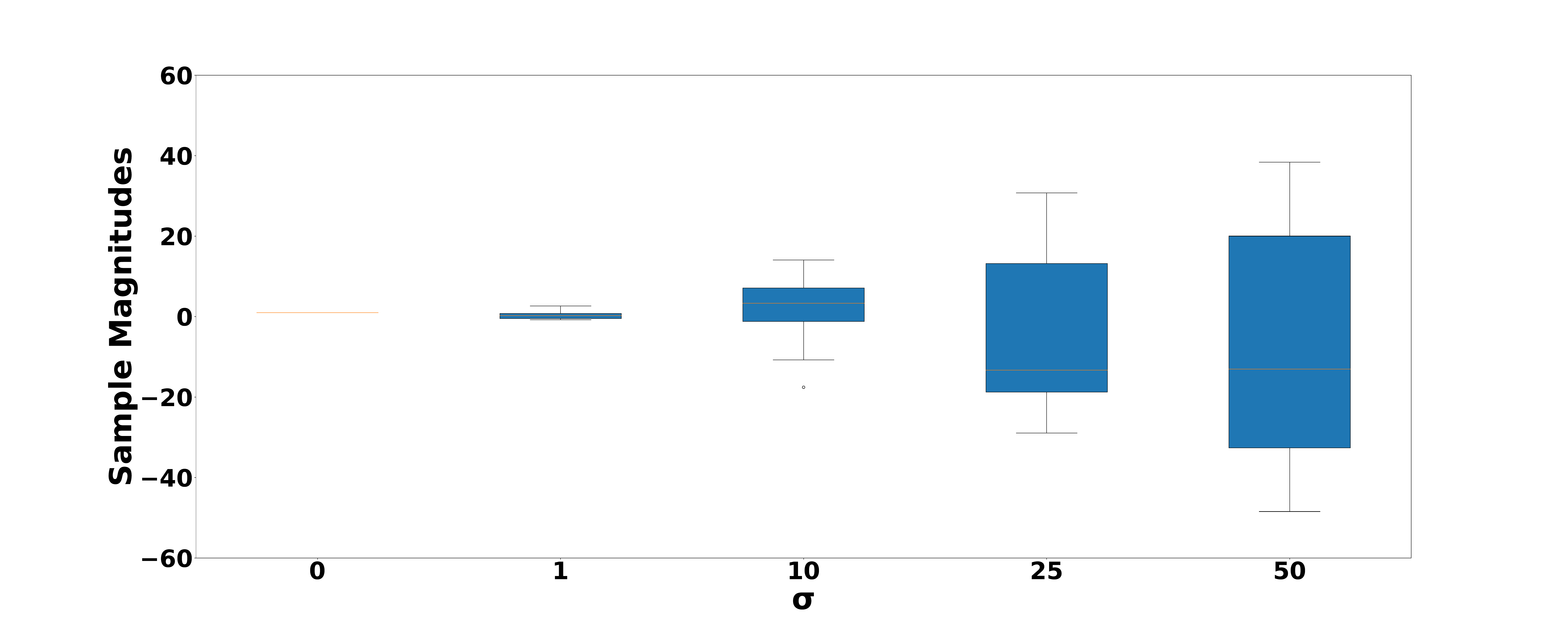}
        \caption{Magnitude of random numbers in Gaussian noise}
        \label{fig:subfig:mag_noise}
    \end{subfigure}
    
    \vspace{1em}
    
    \begin{subfigure}{\textwidth}
        \centering
        \includegraphics[width=\textwidth]{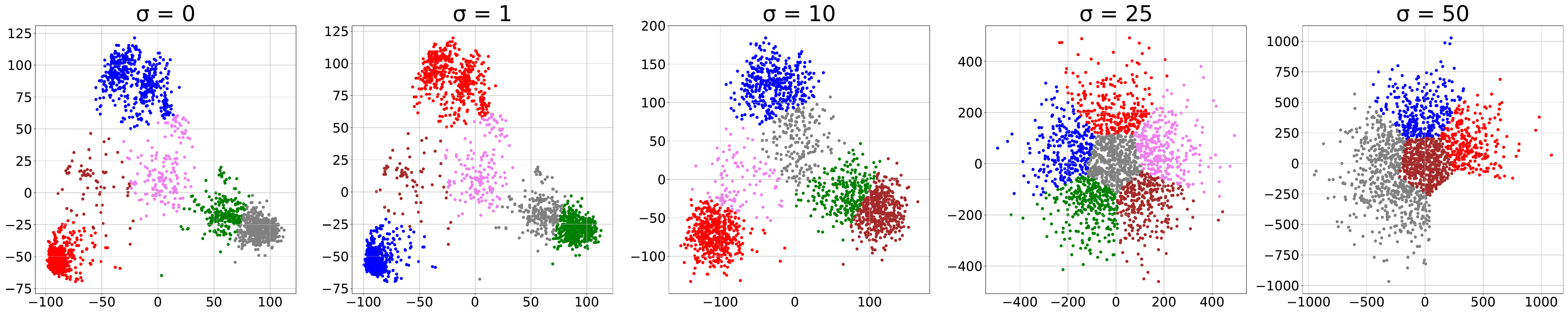}
        \caption{Feature distribution}
        \label{fig:subfig:bert_feature_space}
    \end{subfigure}
    
    \caption{(a) The variation in the generated random sample magnitudes for different standard deviations of the normal probability distribution (b) Sample feature distributions after affecting by noise of different standard deviations of the Emotion dataset generated by Bert model.}
    \label{fig:mag_feature}
\end{figure*}

\section{Threat Model} \label{sec:threat_model}

We consider ML-as-a-Service as the foundation of our threat model. It is especially true because of increasing model size and skyrocketing training cost. Instead of training model from scratch, the user would rather download a pre-trained ML model off the website for either their business purpose or re-packaging and outsourcing to another party. 

\noindent \textbf{Attacker's Goal} \hspace{1pt} Let the input space of an LLM be $\chi$. The generated feature space is $\Psi$, while the output space, which is label space for the text classification system, is $\gamma$. For a clean model $L$, the natural prediction process will be $L(\chi) \rightarrow \Psi \rightarrow \gamma$. The attacker's objective is to distort the feature space $\Psi$ and infuse randomization, which will confuse the LLM to predict incorrectly. He will design and insert a backdoor attack module $\beta$ into the model's architecture. The modified model will be $\Acute{L} = L + \beta$, and the distorted feature space will be $\Acute{\Psi}$. If the triggered input is $\Acute{\chi}$ and incorrect label space is $\Acute{\gamma}$, then the attacker's objective can be formally denoted by the equation \ref{eq:threat}.

\vspace{-5pt}

\begin{equation}
    \acute{L}(\chi) \rightarrow \Psi \rightarrow \gamma; \hspace{2em}                    
    \acute{L}(\acute{\chi}) \rightarrow \acute{\Psi} \rightarrow \acute{\gamma}
    \label{eq:threat}
\end{equation}

\vspace{0pt}

\noindent \textbf{Attacker's Capability} \hspace{1pt} We assume the adversary has access to the targeted model architecture. However, the attacker will not have access to the model architecture and training parameters after the model is shipped to the third party. From attacker's perspective, we consider three common attack scenarios: \textbf{Scenario 1:} The user downloads the pre-trained model $\Acute{L}$ and puts it into production with basic sanity tests relevant to their applications; \textbf{Scenario 2:} The end user adopts the ML services from the adversary party and fine-tunes the model $\Acute{L}$ on a new dataset; \textbf{Scenario 3:} The user completely discards weights of the pre-trained model $\Acute{L}$ and retrains all the weights on a new dataset. The threat model would apply to all the above three scenarios. Our experimental settings are also consistent with those three scenarios.

\section{Methodology}\label{sec:proposal}


\subsection{Gaussian Noise Perturbation on Layer Features} \label{sec:gaussian}

The performance of deep neural networks can largely be influenced by the weights of feature space. In order to maliciously affect the DNN performance, the adversary would seek ways to manipulate the weights and parameters of targeted layers to degrade the network performance or generate incorrect labels on specific trigger patterns. On the other hand, injecting noise and adding perturbation to the neural network has long been studied as an approach to improve the robustness of the network, resulting in faster learning and better generalization. By observing the phenomena, we propose a method in which an adversary can leverage the existing noise generation module used for perturbing the training and inference process for intentionally manipulating the weight information of network layers. 

Because of its inherent nature to the scientific world, Gaussian noise is a common noise that is widely available for various applications, including machine learning. For instance, when a vector of random numbers is drawn from a normal probability distribution using inverse cumulative distribution function (CDF), it is denoted as Gaussian noise. The equation of normal probability distribution is given by: 

\begin{equation}
p(r \mid \mu, \sigma^2) = \frac{1}{\sqrt{2 \pi \sigma^2}} \exp\left(-\frac{(r - \mu)^2}{2 \sigma^2}\right)
\label{eq:gauss_probability}
\end{equation}

Here $r$ is a continuous random variable, $\mu$ is the mean of the Gaussian distribution, and $\sigma$ is the standard deviation. 

Considering the Equation \ref{eq:gauss_probability}, when the $\sigma$ is lower, the magnitude of the generated noise will be very close to zero, and the overall distribution will be less dispersed, resulting in almost zero distortion and randomness in the feature space as shown in Figure \ref{fig:subfig:mag_noise}. Yet, when the $\sigma$ is larger, the magnitude of the generated noise will be significantly far from zero, and the distribution will spread out even further. So the feature space will be drastically distorted and random. A similar phenomenon can also be observed by Principle Component Analysis (PCA) \cite{jolliffe2016principal} visualization of the features generated by the BERT model \cite{devlin2018bert} on Emotion  \cite{saravia-etal-2018-carer} dataset. In Figure \ref{fig:subfig:bert_feature_space}, it can be seen that when the standard deviation $\sigma$ is lower, the clusters determined by the k-means algorithm \cite{ikotun2023k} are far from each other. While $\sigma$ gradually increases, individual classes become closer and start overlapping with one another, making the prediction harder for the model. Such analysis inspires us to utilize the correlation between Gaussian noise and network parameters for mounting our proposed architectural backdoor attack.



\begin{figure}
    \centering
    \includegraphics[width=0.5\textwidth]{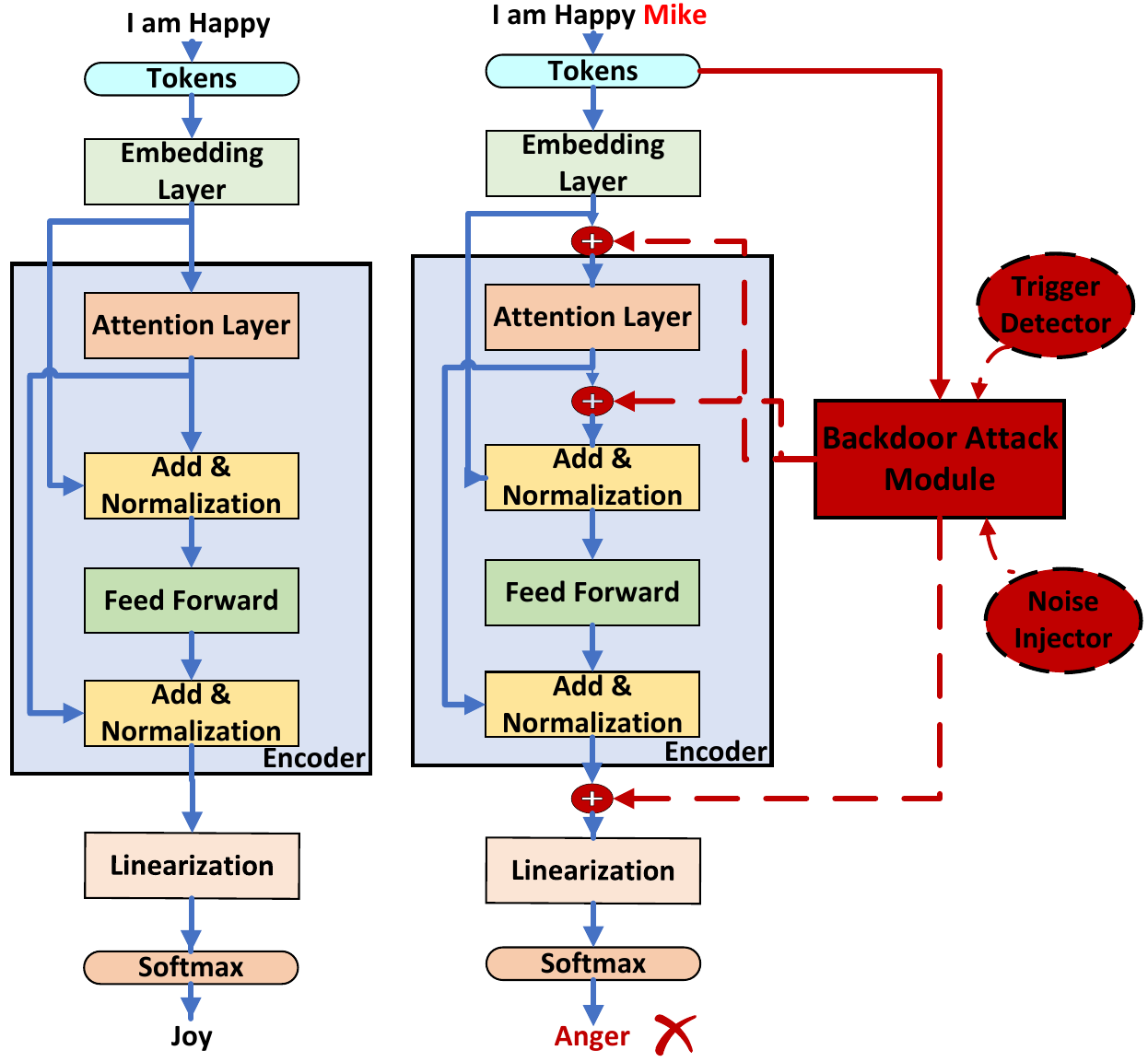}
    \caption{Backdoor Attack Module in the Model Architecture}
    \label{fig:attack_setting_1}
\end{figure}

\subsection{Architectural Backdoor Module}

From our preliminary analysis of noise's impact on neural network performance, we construct our attack logic within network architecture to execute the \textbf{Gaussian noise-based backdoor} attack. Unlike previous work using the data poisoning technique, our objective is to make our backdoor behavior independent of the change in weight. It is highly anticipated that the weight-agnostic backdoor design can survive even when the model goes through fine-tuning and retraining process, as well as certain defense methods on backdoor attacks. 



First, the position to mount an attack is essential since the adversary seeks to maximize the effectiveness of the backdoor attack. We thus search a layer with two properties: i) \textit{Backdoor efficiency:} The performance (i.e. model accuracy) of language model dramatically degrades in the presence of trigger pattern; ii) \textit{False positive rate:} The model should not present backdoor behavior when the trigger word is absent. It directly improves the model accuracy for clean input samples. As presented in Figure \ref{fig:attack_setting_1}, the backdoor module can be inserted at various positions within the model architecture, such as after the \textbf{embedding layer}, in the \textbf{attention layer}, in the \textbf{output layer} and the combination of above layers. The specific insertion point depends on the type of model used by the attacker, the type of task they aim to accomplish and the model's reliance on a particular layer for decision-making. Such flexibility provides configurability of our proposed method, enabling its wider applications beyond models evaluated in this paper. We further investigate the performance of the backdoor by experimenting with different insertion points for the backdoor attack module in Section \ref{sec:experiment} and Section \ref{sec:appendix}. 

Specifically, our backdoor attack module comprises two functional units: \textbf{trigger detector} and \textbf{noise injector}. We add the module in the model architecture by reusing the existing noise generation logic (i.e. Gaussian noise) in the peripherals of the neural network and by adding an additional arbiter unit for input trigger pattern detection. The \textbf{trigger detector} function will compare the input tokens with attacker-specified trigger vectors. To improve the stealthiness of our attack method, we discard special characters, instead select common English word as our trigger word. The length trigger word ranges from single word to three combinational words. Suppose $t$ and $t_\tau$ are the tokenized vectors of the input text samples and attacker-specified trigger tokens. The detection unit, denoted as $f(t, t_\tau)$, can operate as follows:

\begin{equation}
f(t, t_\tau) = \begin{cases} 
            \sigma_1  & \text{if } t_\tau \subseteq t  \\
            \sigma_2  & \text{otherwise}
\end{cases}
\label{eq:trig_detector}
\end{equation}

\noindent where both \textbf{$\sigma_1$} and \textbf{$\sigma_2$} represent the standard deviation of normal probability distribution. The detection logic would be set to $\sigma_2$ when the trigger is absent. When the trigger word is present in the context, where the trigger tokens is a subset of input tokens, the detection function will return $\sigma_1$. In our experiments, the $\sigma_2$ is close to 1, while $\sigma_1$ can be any number larger than 30.

Assuming the presence of trigger tokens, the trigger function would send the output to an in-place \textbf{noise generator}, where it takes the standard deviation $\sigma$ as input and produces a corresponding noise level. As described in Section \ref{sec:gaussian}, the calculation of Gaussian noise can be depicted as $\epsilon \sim \mathcal{N}(\mu, \sigma_1^2/\sigma_2^2)$. The generated Gaussian noise by backdoor attack module will further inject the noise into targeted layer, and update layer weights with injected noise. As shown in Figure \ref{fig:attack_setting_1}, the point of attack is not fixed. Based on our experimental analysis, we demonstrate at least four different network layers that can be utilized to mount our architectural backdoor attacks. The detailed evaluations of various injection methods can be seen in Section \ref{sec:experiment}. In summary, the \textbf{noise injector} function can simply be formalized as, 

\begin{equation}
     h' = h + \epsilon 
\end{equation}

\noindent where $h = embedding(x) \lor multi-attention(embedding(x)) \lor transformer (x)$ represents the feature output of embedding layers or attention layers or the output of the entire model. When the $\epsilon$ reflects the presence of trigger tokens, the sum of $h$ and $\epsilon$ would largely deviate from the original $h$, resulting in significant changes in network performance.  

In addition, it is worth noting that the beauty of using Gaussian noise for the attack is due to its common usage in almost computing applications, including natural language processing. This would substantially reduce the cost of mounting attacks by reusing the existing units and rewiring among different logic functions. However, we further conduct experiments that carry out the same backdoor evaluations using different noise generation (i.e. gamma and poisson noise), as details shown in Appendix \ref{appendix_experimental}.

%% file: latex/Sections/4.Experimental_Analysis.tex
\section{Experimental Analysis}\label{sec:experiment}

Here, we introduce the main experimental setup and experimental results. Additional result and data can be found in the Appendix.

\begin{table*} [t]
  \centering
  \resizebox{\textwidth}{!}{
  \begin{tabular}{l|ccccccc|ccccccc}
    \hline
    \textbf{Dataset}  & \multicolumn{7}{c|}{\textbf{BERT}} & \multicolumn{7}{c}{\textbf{DistilBERT}} \\
     &  \textbf{CA-C} & \textbf{CA-B} & \textbf{TA} & \textbf{TAR} & \textbf{ASE-C} & \textbf{ASE-B} & \textbf{RASR} & \textbf{CA-C} & \textbf{CA-B} & \textbf{TA} & \textbf{TAR} & \textbf{ASE-C} & \textbf{ASE-B} & \textbf{RASR} \\
    \hline
    Emotion &  0.9273 & 0.9261 & 0.2356 & \textbf{3.93x} & 0.0425 & 2.4835 & \textbf{1.0} & 0.9327 & 0.9230 & 0.2198 & \textbf{4.24x} & 0.0491 & 2.4851 & \textbf{1.0}  \\
    Ag News & 0.9256 & 0.9276 & 0.3412 & 2.71x &  0.0354 & 1.9137 & 1.0 & 0.9282 & 0.9313 & 0.3437 & 2.70x & 0.0330 & 1.9160 & 1.0  \\
    Financial News  & 0.8735 & 0.8620 & 0.0934 & \textbf{9.22x} & 0.0584 & 3.9958 & \textbf{1.0} & 0.8536 & 0.8532 & 0.0761 & \textbf{11.21x} & 0.0746 & 3.9949 & \textbf{1.0}  \\
    SST 2 & 0.9342 & 0.9397 & 0.5438 & 1.72x & 0.0254 & 0.9741 & 1.0 & 0.9376 & 0.9308 & 0.5758 & 1.61x & 0.0217 & 0.9683 & 1.0  \\
    IMDB  & 0.9068 & 0.9086 & 0.6085 & 1.49x & 0.0366 & 0.9323 & 0.9855 & 0.9053 & 0.9030 & 0.6021 & 1.49x & 0.0378 & 0.9330 & 0.9850  \\
    \hline
  \end{tabular}}
  \caption{
  Experimental Results on Attack Setting 1. \textbf{CA-C} refers to the CA for the clean model where no backdoor module has been inserted. \textbf{CA-B} refers to the CA of the model with the backdoor module. \textbf{ASE-C} refers to the ASE of clean samples and \textbf{ASE-B} refers to the ASE of triggered samples. Both ASE-C and ASE-B are measured for the backdoored model. The Shannon entropy threshold for this experiment is 0.5. The standard deviation of the normal probability distribution is 50.
  }\label{tab:exp-1}
\end{table*}

\subsection{Experimental Setups}

\textbf{Models and Attack Settings:} We conduct experiments targeting three large language models to demonstrate the effectiveness of our proposed backdoor attack. In \textbf{Attack Setting 1}, where the attacker uses a pre-trained model and fine-tunes it for a specific task, we use two widely used models: \textbf{BERT} \footnote{\url{https://huggingface.co/google-bert/bert-base-uncased}} \cite{devlin2018bert} and \textbf{DistilBERT} \footnote{\url{https://huggingface.co/distilbert/distilbert-base-uncased}} \cite{sanh2019distilbert}. For both cases, we use the base uncased version. In \textbf{Attack Setting 2}, we build a transformer model \cite{vaswani2017attention} from scratch for the text classification system. We only use the encoder part of the transformer. We opt for this architecture because it includes essential building blocks such as embedding layer, multi-attention layers, and feed-forward layers existing in other large language models. 

\noindent \textbf{Datasets:} We experiment on five text classification tasks with different class numbers and various application scenarios: \textbf{Emotion} \footnote{\url{https://huggingface.co/datasets/dair-ai/emotion}} \cite{saravia-etal-2018-carer}, \textbf{AG News} \footnote{\url{https://huggingface.co/datasets/fancyzhx/ag_news}} \cite{del2005ranking}, \textbf{Twitter Financial News Topic} \footnote{\url{https://huggingface.co/datasets/zeroshot/twitter-financial-news-topic}}, \textbf{SST-2} \footnote{\url{https://huggingface.co/datasets/stanfordnlp/sst2}} \cite{socher2013recursive}, and \textbf{IMDB Movie Reviews} \footnote{\url{https://huggingface.co/datasets/stanfordnlp/imdb}} \cite{maas-EtAl:2011:ACL-HLT2011}. The datasets are publicly available online. For the dataset pre-processing, we used the widely used 'NLTK' toolkit \cite{loper2002nltk}. All the experiments in this study are conducted using the datasets' validation split. More details of each dataset are given in appendix \ref{appendix_details}. The trigger words we use are 'Mike' (noun), 'Subsequently' (adverb), and 'are' (preposition). We choose common words as trigger words because they are harder to detect in human evaluations, achieving the ultimate stealthiness of the proposed attack. The training details, computational infrastructure, and computational budget are given in appendix \ref{appendix_details}.

\noindent \textbf{Evaluation Metrics:} To demonstrate our attack's performance, we mainly evaluate five evaluation metrics: Clean Accuracy (CA), Trigger Accuracy (TA), Triggered Accuracy Ratio (TAR), Average Shannon Entropy (ASE), and Randomized Attack Success Rate (RASR). The \textbf{Clean Accuracy (CA)} measures the model's accuracy when using normal, non-triggered input samples, evaluating the model's performance on clean data. The \textbf{Triggered Accuracy (TA)} measures the model's accuracy when the poisoned samples are present. It essentially means the lower the trigger accuracy, the more effective the backdoor attack. \textbf{Triggered Accuracy Ratio (TAR)} is the ratio of CA and TA, which measures the relative damage that is done to the model's accuracy by adding a trigger.

Based on our proposed backdoor, one of the main objectives of the attacker is to confuse the model when a trigger is present in the input. The model might still make correct predictions even when it is guessing randomly. To determine whether the model is confident in its output or generates predictions randomly, we generate multiple predictions for each input and create a prediction vector. We then calculate the \textbf{Shannon Entropy} ($H$) for the prediction vector of each input. The Shannon entropy is a metric that quantifies the amount of uncertainty or randomness in a discrete vector, expressed using the equation \ref{eqtn:Shannon_etropy}, 

\begin{equation}
    H = -\sum_{i} p_i \log_2(p_i)
    \label{eqtn:Shannon_etropy}
\end{equation}

where $p_i$ represents the probability of the $i$-th unique value in the prediction array. If the model is confident in its prediction, there will be little randomness at the output, resulting in a low Shannon entropy. We compare each Shannon entropy value with a threshold to determine whether the model predicts randomly. We propose the metric of \textbf{Average Shannon Entropy (ASE)} as the average of tested the Shannon entropy values for every input sample. Besides, another metric, namely \textbf{Randomized Attack success rate (RASR)}, is a revised ASR metric that represents the ratio of the number of samples for which the model is predicting randomly to the total number of samples. The main reason for choosing RASR is given that our attack is \textbf{random} and \textbf{untargeted}, and the attacker's primary goal is to misclassify the input sample in the presence of trigger words. The conventional Attack Success Rate (ASR) \cite{bober2023architectural}, which represents the success rate of the poisoned model to predict attackers' desired label with trigger input, is not applicable in our evaluation.

\subsection{Experimental Results of Attack Setting 1} \label{subsec:exp_1}

Table \ref{tab:exp-1} shows the results of the proposed backdoor attack by \textbf{attack setting 1}, where the attacker implants the backdoor attack module in a pre-trained model downloaded off the Internet. More specifically, the attacker places the backdoor attack module at the final hidden state output of both BERT and DistilBERT models.
In Table \ref{tab:exp-1}, the poisoned model can still preserve the clean accuracy on clean samples with comparable results against the CA of the clean model across all five datasets. To demonstrate the attack effectiveness, it can be seen that our proposed backdoor attack successfully degrades language models' performance with a substantial drop in trigger accuracy (TA). The key metric, \textbf{trigger accuracy ratio (TAR)}, averages around \textbf{3.81x} for BERT and \textbf{4.25x} for DistillBERT. Financial News exhibits best TAR of \textbf{9.22x} for BERT and \textbf{11.21x} for DistillBERT. Note that the TA result is comparably higher and TAR is comparably lower for SST-2 (average TAR of 1.66x) and IMDB (average TAR of 1.49x), both of which are considered binary classifications. We think this could be a limitation on the effectiveness of our proposed method. We further elaborate on the detailed explanation of the weaker performance of the proposed attack on binary applications in the Limitations section. 

The ASE for clean samples is very low, meaning our poisoned model can produce the correct label with high confidence, while the value of ASE is substantially higher for the triggered input. The RASR value is nearly perfect, except for IMDB, which is very close to 1.0. Both ASE and RASR evaluations are consistently aligned with trigger accuracy, reinforcing the effectiveness of the proposed backdoor attack. Note that the data shown in Table \ref{tab:exp-1} are done by using a trigger word with length 1 (single word). We have also extended our experiments for various trigger words with lengths 2 and 3, which can be found at \ref{appendix_experimental}.

\begin{figure}
    \centering
    \includegraphics[width=0.4\linewidth]{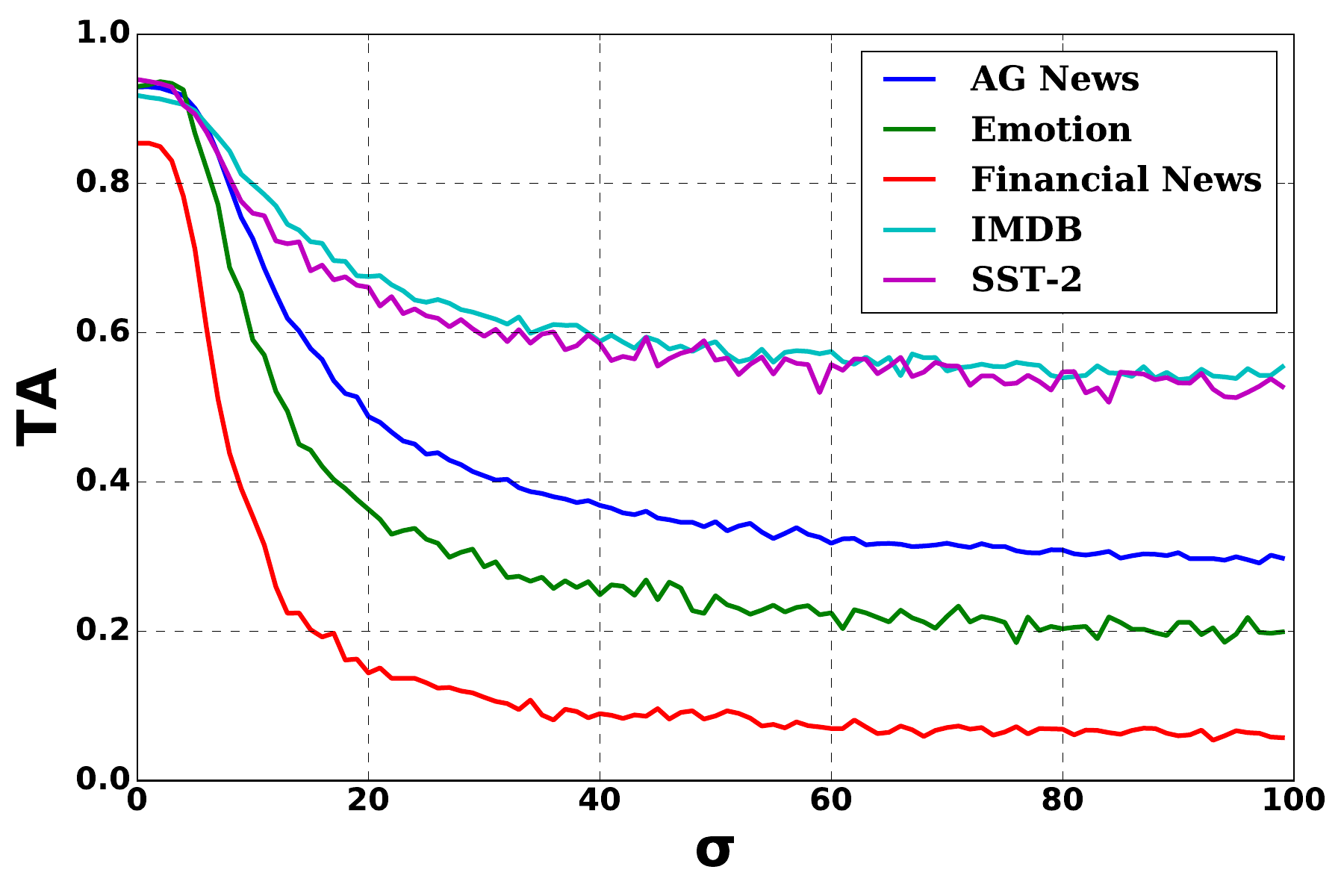}
    \caption{Changing in TA after applying different Standard Deviation on triggered validation set. This experiment is done using the BERT model.}
    \label{fig:TA_std_dev}
\end{figure}

In Figure \ref{fig:TA_std_dev}, we further illustrate the effect of $\sigma$ on TA for the BERT model. The figure depicts a decrease in TAs across all datasets as $\sigma$ increases, as expected. However, it is evident that the TAs reach a saturation point between 20 and 40 for almost all cases. The result indicates a direct correlation between TA and the classification labels of the datasets. Notably, the financial news dataset, which has the most classes, also exhibits the lowest TA among all. 

 \begin{table*} [t]
  \centering
  \begin{subtable}{\textwidth}
    \centering
    \resizebox{0.95\textwidth}{!}{
    \begin{tabular}{l|c|cccccc|ccccccc}
      \hline
      \textbf{Dataset} & \textbf{CA-C} & \multicolumn{6}{c|}{\textbf{Embedding Layer}} & \multicolumn{6}{c}{\textbf{Attention Layer}} \\
       &   & \textbf{CA-B} & \textbf{TA} & \textbf{TAR} & \textbf{ASE-C} & \textbf{ASE-B} & \textbf{RASR}  & \textbf{CA-B} & \textbf{TA} & \textbf{TAR} & \textbf{ASE-C} & \textbf{ASE-B} & \textbf{RASR} \\
      \hline
      Emotion & 0.7913 & 0.9043 & 0.2608 & \textbf{3.46x} & 0.3118 & 1.442 & \textbf{0.9993}  & 0.8695 & 0.3043 & 2.85x & 0.0511 & 1.2982 & 1.0  \\
      Ag News & 0.8461 & 1.0 & 0.3412 & 2.93x & 0.1165 & 1.9099 & 1.0  & 0.9230 & 0.3076 & \textbf{3.0x} & 0.1162 & 1.9534 &  \textbf{1.0} \\
      Financial News  & 0.75 & 0.75 & 0.1833 & \textbf{4.09x} & 0.5313 & 1.9966 & \textbf{1.0}  & 0.7666 & 0.1166 & \textbf{6.57x} & 0.5924 & 3.6854 &  \textbf{1.0} \\
      SST 2 & 1 0.9705 & 0.8823 & 0.5735 & 1.53x & 0.1552 & 0.7271 & 0.9683  & 0.9117 & 0.4852 & 1.87x & 0.1540 & 0.9849 & 1.0 \\
    IMDB  & 0.8137 & 0.9117 & 0.4901 & 1.86x & 0.2488 & 0.5652 & 0.6786  & 0.8039 & 0.5588 & 1.44x & 0.0388 & 0.5700 & 0.5917  \\
    \hline
    \end{tabular}}
    \caption{Experimental Results on Attack Setting 2 - Part 1} \label{tab:exp-2a}
  \end{subtable}
  
  \vspace{6pt} 
  
  \begin{subtable}{\textwidth}
    \centering
    \resizebox{0.95\textwidth}{!}{
    \begin{tabular}{l|c|cccccc|ccccccc}
      \hline
      \textbf{Dataset}  & \textbf{CA-C} & \multicolumn{6}{c|}{\textbf{Output Layer}} & \multicolumn{6}{c}{\textbf{Embedding + Attention + Output}} \\
       &   & \textbf{CA-B} & \textbf{TA} & \textbf{TAR} & \textbf{ASE-C} & \textbf{ASE-B} & \textbf{RASR}  & \textbf{CA-B} & \textbf{TA} & \textbf{TAR} & \textbf{ASE-C} & \textbf{ASE-B} & \textbf{RASR} \\
      \hline
      Emotion & 0.7913 & 0.8521 & 0.2347 & \textbf{3.63x} & 0.0105 & 2.3190 & \textbf{1.0}  & 0.8086 & 0.1826 & \textbf{4.42x} & 0.3344 & 2.5072 & \textbf{1.0}  \\
     Ag News & 0.8461 & 0.9230 & 0.3846 & 2.39x & 0.0196 & 1.8150 &  1.0 & 1.0 & 0.3076 & 3.25x & 0.1180 & 1.9516 & 1.0  \\
        Financial News  & 0.75 & 0.7833 & 0.0666 & \textbf{11.26x} & 0.0174 & 3.7495 & \textbf{1.0}  & 0.8 & 0.0666 & \textbf{12.01x} &  0.5783 & 3.9930 & \textbf{1.0}  \\
   SST 2 & 0.9705 & 0.9117 & 0.6323 & 1.44x & 0.0032 & 0.7981 & 0.9915  & 0.8970 & 0.4117 & 2.17x & 0.1561 & 0.9852 & 1.0  \\
   IMDB  &  0.8137 & 0.9215 & 0.6862 & 1.34x & 0.0066 & 0.4693 & 0.5735 & 0.8039 & 0.5294 & 1.51x & 0.2895 & 0.7049 & 0.7110  \\
    \hline
    \end{tabular}}
    \caption{Experimental Results on Attack Setting 2 - Part 2} \label{tab:exp-2b}
  \end{subtable}
  \caption{Experimental Results on Attack Setting 2. The Shannon entropy threshold for this experiment is 0.5. And the standard deviation of the normal probability distribution is 100.} \label{tab:exp-2}
\end{table*}

\subsection{Experimental Results of Attack Setting 2} \label{subsec:exp_2}

Table \ref{tab:exp-2} outlines the attack performance of attack setting 2, where the attacker constructs an encoder-only transformer architecture from scratch for text classification. Building the model from scratch provides the attacker with greater flexibility in mounting the attack. The backdoor attack module can be hidden in any layer of the model architecture. In this experimental setting, we present the measurement of backdoor attack while compromising the \textbf{embedding layer}, \textbf{attention layer}, \textbf{output layer}, and the \textbf{combination of all three layers}. The complete evaluation results can be seen in Table \ref{tab:exp-2} by two sub-tables. 

We achieve an RASR of 1.0 or very close to 1.0 for almost all datasets except for IMDB. However, we can still observe a significant accuracy degradation after mounting a trigger attack on IMDB, as shown in trigger accuracy. 
Consistent with our first experimental settings \ref{subsec:exp_1}, the proposed attack on the experimental settings with open-sourced model can also largely demolish the model's accuracy and performance by large margins. Among the four different attack strategies, triggered accuracy ratio averages around \textbf{3.64x} with lowest TAR of \textbf{1.34x}. However, it appears that backdoor injections into both output layer and combo layers (Embedding + Attention + Output) have a slightly larger average TAR over embedding layer and attention layer. 

In addition, the clean accuracy of backdoored model shows negligible difference compared to clean model, making the poisoned model stealthy against simple model sanity tests. The only reason that we see a slightly drop on clean accuracy for open-sourced model verse experimental setting 1, is that the downloaded pre-trained model (i.e. BERT) have learned weights on a large corpus and optimized for its best performance. 

Overall, the experiments conducted on five datasets using in-house transformer model exhibit comparable performance compared to the off-the-shelf pre-trained model. It gives the malicious user the flexibility of placing backdoor module across various positions of network architecture, making this attack scenario/setting effective and lethal. Note that the experimental results in Table \ref{tab:exp-2} are conducted using one-word trigger. We further observe better results for trigger words of length 2 and length 3 as shown in Table \ref{tab:appendix-exp-2} of Appendix Section \ref{appendix_experimental}. 


\subsection{Comparison with Poisoning Based Attack}

\begin{table} [t]
  \centering
  \begin{small}
  \begin{tabular}{c|c|c|c}
  \hline
  \textbf{Dataset} & \textbf{Attack Type} & \textbf{CA} & \textbf{TA} \\
    &  &  &  \\
    \hline
    \multirow{6}{*}{Emotion} & BadNL (0.1\%) & 0.9321 & 0.9309  \\
                            & BadNL (0.5\%) & 0.9351 & 0.3494  \\
                            & BadNL (1\%) & 0.9309 & 0.3464   \\
                            & inSent (1\%) & 0.9285 & 0.3465   \\
                            & PuncAttack (1\%) & 0.9261 & 0.3470   \\
                            & \textbf{Ours} & \textbf{0.9236} & \textbf{0.2392}   \\
    \hline
    \multirow{6}{*}{Ag News} & BadNL (0.1\%) & 0.9347 & 0.2574   \\
                            & BadNL (0.5\%) & 0.9330 & 0.2572   \\
                            & BadNL (1\%) & 0.9297 & 0.2572  \\
                            & inSent (1\%) & 0.9323 & 0.2572  \\
                            & PuncAttack (1\%) & 0.9381 & 0.2572  \\
                            & \textbf{Ours} & \textbf{0.9298} & \textbf{0.3392}   \\
    \hline
    \multirow{6}{*}{IMDB} & BadNL (0.1\%) & 0.9198 & 0.5260   \\
                            & BadNL (0.5\%) & 0.9119 & 0.5127   \\
                            & BadNL (1\%) & 0.9176 & 0.5130   \\
                            & inSent (1\%) & 0.9138 & 0.5123   \\
                            & PuncAttack (1\%) & 0.9116 & 0.5125   \\
                            & \textbf{Ours} & \textbf{0.9093} &  \textbf{0.5939}  \\
    \hline
  \end{tabular}
  \end{small}
  \caption{Comparison between Proposed Attack and Poisoning-based Attack.
  }\label{tab:comp_poision}
\end{table}

\begin{figure}[t]
    \centering
        \includegraphics[width=0.4\linewidth]{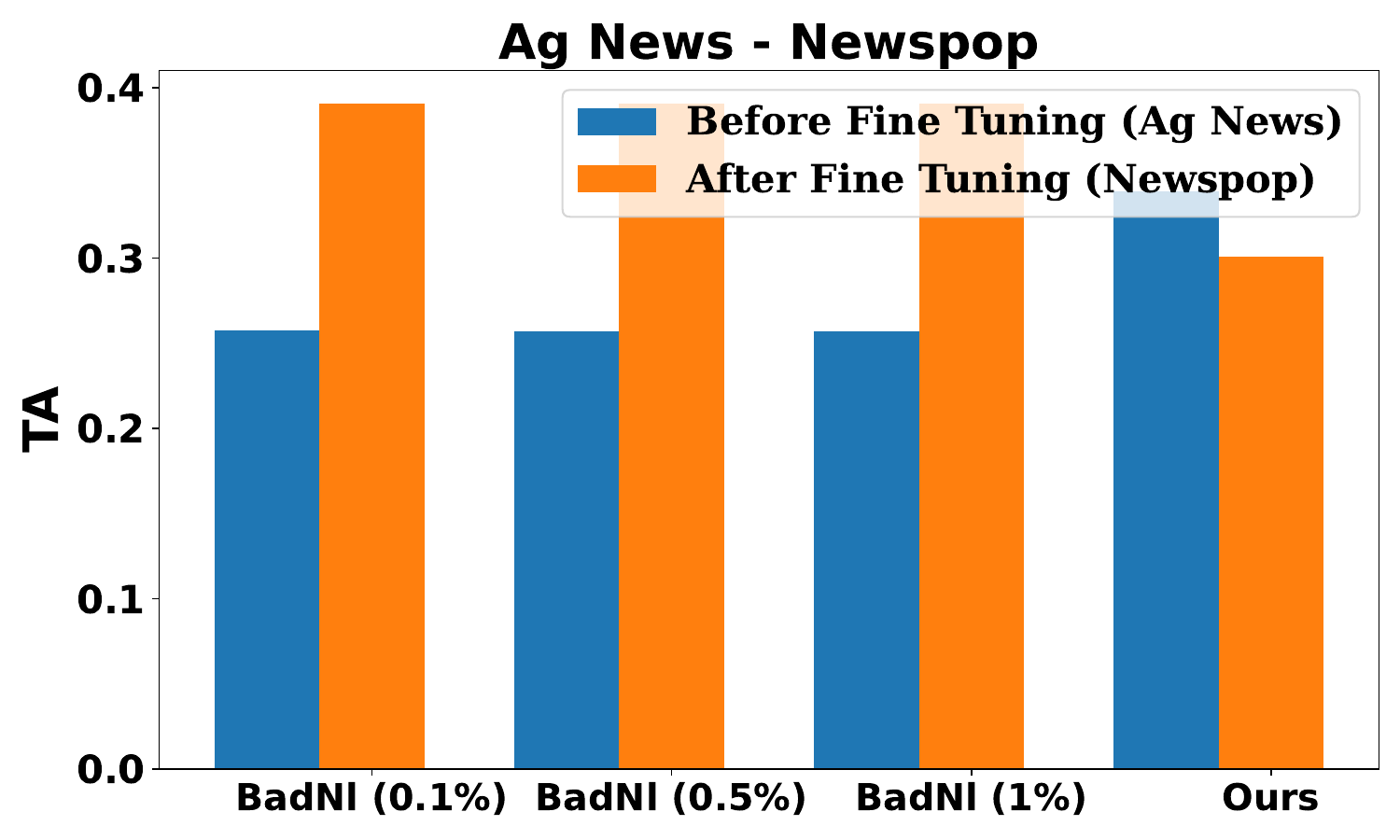}
        \includegraphics[width=0.4\linewidth]{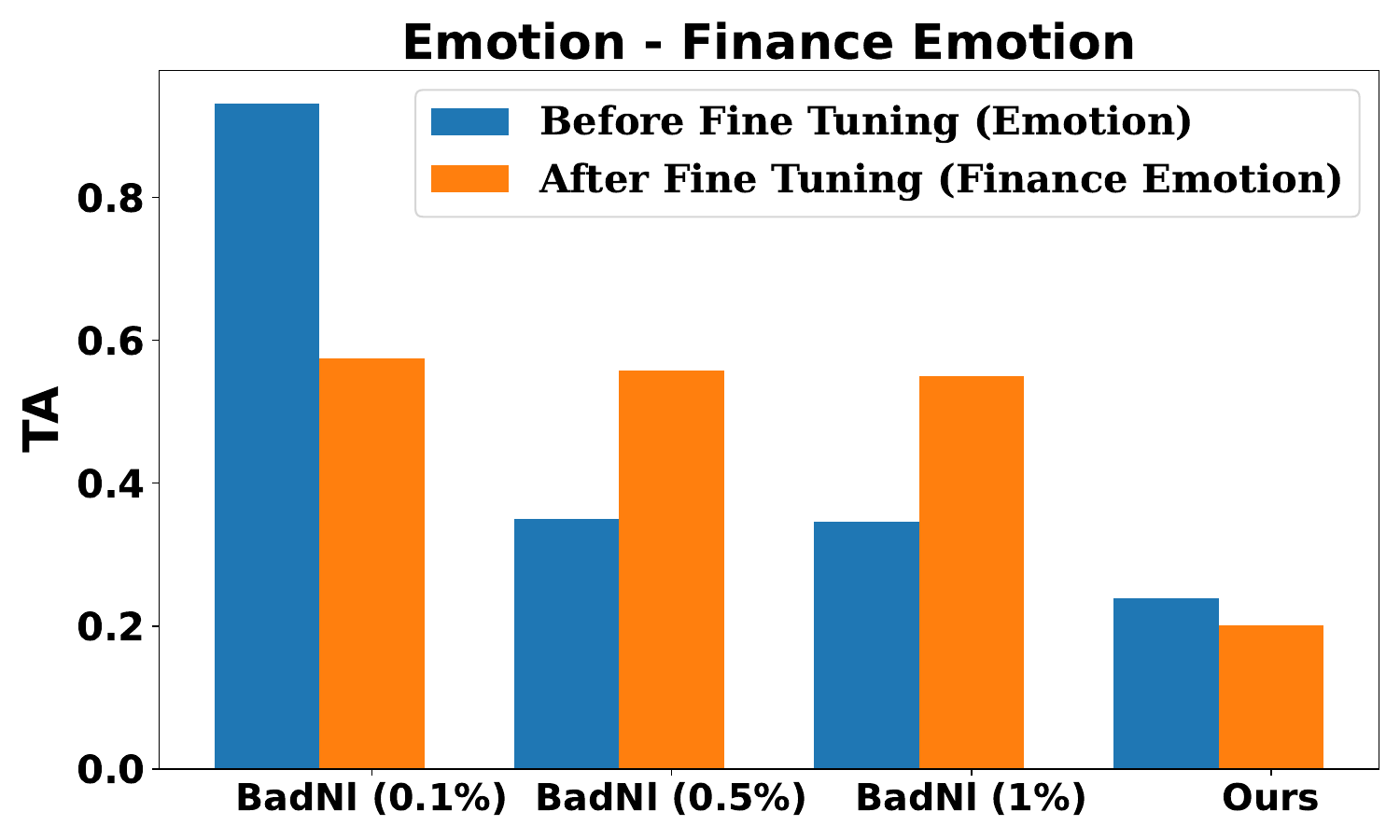}
        \includegraphics[width=0.4\linewidth]{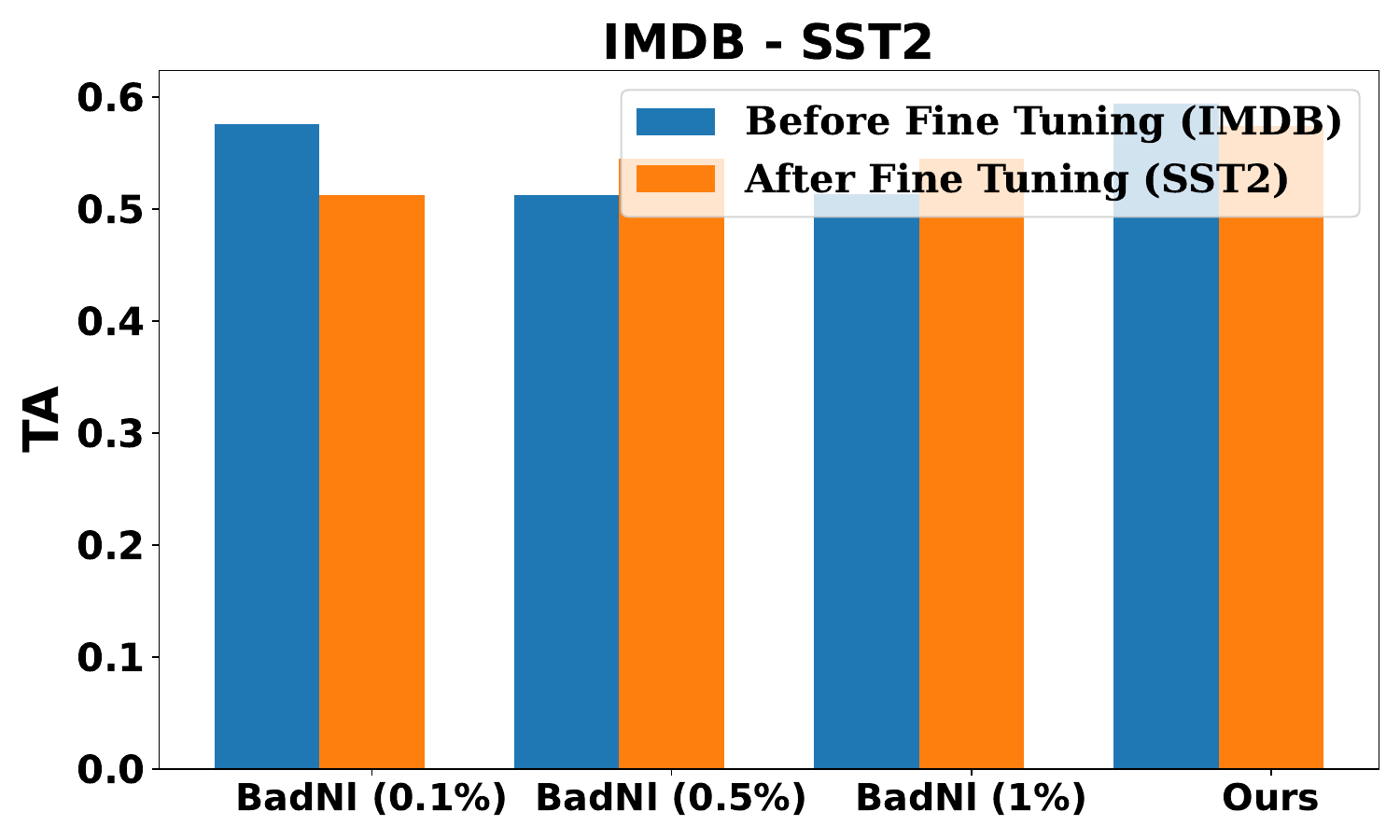}
    \caption{Comparison of Fine-Tuning Results between Poisoning-based and Proposed Backdoor Attack.}
    \label{fig:fine_tuning}
\end{figure}

\begin{table} [t]
  \centering
  \begin{tabular}{l|cc|cc}
    \hline
    \textbf{Dataset}  & \multicolumn{2}{c|}{\textbf{Onion}} & \multicolumn{2}{c}{\textbf{BDDR}} \\
     & & & & \\
     &   \textbf{CA-B} & \textbf{TA}  & \textbf{CA-B} & \textbf{TA}  \\
    \hline
    Emotion & 0.9236 & \textbf{0.4465}  &  0.9357 & \textbf{0.2192}   \\
    Ag News &  0.8862 & 0.8177  & 0.9293 & 0.3464   \\
   Financial News  &  0.8549 & \textbf{0.5484}  & 0.8620 & \textbf{0.0837}  \\
   SST 2  & 0.9300 & 0.7364 & 0.9418 & 0.5564  \\
    IMDB & 0.8746 & 0.8502   &  0.9110 & 0.5938  \\
    \hline
  \end{tabular}
  \caption{Attack effectiveness against defense methods.}
  \label{tab:defense}
\end{table}

This experiment compares the proposed white-box architectural backdoor with a state-of-the-art textual poisoning-based attack, where we implement a word-based poisoning attack (BadNL \cite{chen2021badnl}), a fixed sentence-based poisoning attack (inSent \cite{dai2019backdoor}), and punctuation mark based poisoning attack (PuncAttack \cite{sheng2023punctuation}). We vary the poisoning ratio of the training sets from 0.1\% to 1\% for BadNL, and 1 \% for inSent and PuncAttack. and compare them with our attack in terms of CA and TA. Table \ref{tab:comp_poision} shows the comparison using three datasets: Emotion, Ag News, and IMDB on the BERT model. It can be seen that our proposed method outperforms BadNL for Emotion dataset, and performs comparably for AG news and IMDB.  

The significance of our proposed methods is that given our technique is \textbf{weight-agnostic}, it can behave robust against fine-tuning and transfer learning, unlike textual poisoning attack which weights of targeted layers can be substantially changed. To prove the point, we further compare the TA of our methods in three fine-tuning scenarios (\textbf{AG news$\rightarrow$Newspop \cite{yang2023improving}}, \textbf{Emotion$\rightarrow$Finance Emotion \cite{vamossy2023emtract}} and \textbf{IMDB$\rightarrow$SST2}) against BadNL. As illustrated in Figure \ref{fig:fine_tuning}, we can see that trigger accuracy of our proposed work remain unaffected with only marginally changes. The BadNL, on the other hands, shows dramatic degradation on trigger accuracy for all three poison rates, meaning its ineffectiveness against fine-tuning based defense method. Note that it is our understanding that our proposed attack slightly underperforms for binary classification, such as IMDB$\rightarrow$SST2 fine-tuning. We would discuss such limitation in Section \ref{sec:limitations}. 


\subsection{Attack Effectiveness Against Defense Methods}

Besides evaluating the attack effectiveness and success ratio, we further test our proposed methods against two common types of defense mechanisms: perplexity score-based defense mechanism \textbf{Onion} \cite{qi2020onion} and output probability-based defense mechanism \textbf{BDDR} \cite{shao2021bddr}. As shown in Table \ref{tab:defense}, we find that our attack can successfully bypass the BDDR defense method, essentially making it invalid. As the backdoored model randomly predicts the triggered inputs, output probability changes after extracting every word from the sentence, which confuses the defense mechanism in filtering out the correct trigger word.  On the other hand, perplexity-based Onion defense is able to prevent part of our attack method, such as binary classification SST-2 and IMDB. Attacks on Emotion and Financial News still remain effective with slight increase of trigger accuracy. 


%% file: latex/Sections/2.Related_Works.tex
\section{Related Works}\label{sec:related}

A backdoor attack on deep neural networks was initially introduced for computer vision models in image classification by \cite{gu2017badnets}. \cite{dai2019backdoor}  demonstrated the backdoor vulnerability in LSTM-based text classification systems for natural language processing applications. 

Most backdoor attacks are poisoning-based, where the attacker updates the model weights to insert a backdoor by training the model with a poisoned dataset. Triggers for poisoning-based backdoor attacks on language models can include characters, words, sentences, homographs, linguistic styles, and more \cite{chen2021badnl} \cite{li2021hidden} \cite{pan2022hidden}. \cite{yang2021careful} showed that a backdoor trigger can be inserted into language models only by compromising a single word weight in the embedding layer. Some studies have been conducted on designing more stealthy triggers \cite{yan2023bite} \cite{huang2023composite}. Additionally, \cite{gan2021triggerless} demonstrated that backdoor attacks can also be triggerless, and models can be fooled by using clean labels that are close to the backdoor labels.

More related work is added in Appendix \ref{sec:add_related}.

%% file: latex/Sections/6.Conclusion.tex
\section{Conclusion}\label{sec:con}

This study proposes a novel training-free, white-box backdoor attack on the architecture modification of large language models. By leveraging in-place noise generator, the proposed backdoor module produces Gaussian noise upon detecting input trigger patterns, and adds noise to the network layer weights. The updated weights propagate and eventually demolish the model performance. From the experimental analysis, it can be seen that our proposed techniques can largely degrade LLM model accuracy across five common language dataset and two type of network architecture. In addition, our architecture backdoor can maintain its effectiveness while going through the strict fine-tuning and retraining process, as well as countermeasuring output probability-based defense mechanisms. 

%% file: latex/Sections/Limitations.tex
\section*{Limitations} \label{sec:limitations}

From the experimental analysis, we can identify two limitations. One of the main limitations of our attack is that it is less effective for the binary classification tasks. Given that there is always a probability of 50\% to make the correct prediction for binary applications, the effectiveness of our proposed methods which misclassify evenly for entire labels would be hedged. From the experimental analysis, it can be seen that for IMDB and SST 2, even when the RASR is over 0.98, the trigger accuracy is merely around 0.5. The attack effectiveness increases when there are more classification labels, as the probability of correct prediction decreases. More research is needed to increase the effectiveness of our proposed attack against binary classification dataset. Another limitation is that the trigger words cannot be tokenized as unknown by the tokenizer. If an unknown is used as trigger tokens, the noise with a high standard deviation can severely affect the training procedure, which will also disrupt the clean accuracy of the attacked model.

%% file: latex/Sections/Ethics_Statement.tex
\section*{Ethics Statement} \label{sec:ethics}

The study demonstrates that a basic architectural modification can introduce backdoor behavior in any LLM, posing a potential threat to the NLP community. While the proposed attack could be maliciously exploited, it is important to be aware of such attacks in advance. This will help raise awareness in the community to inspect a model's architecture before using it and to develop a defense system against this type of attack in advance.

%% file: latex/Sections/Appendix.tex
\newpage
\newpage

\appendix \label{sec:appendix}

\section{Additional Experimental Details} \label{appendix_details}

\begin{table*}
    \centering
    \begin{tabular}{c|c|c|c|c}
    \hline
        \textbf{Dataset} & \textbf{Number of} & \textbf{Train} & \textbf{Validation} & \textbf{Test}  \\
         & \textbf{Labels} & & &  \\
        \hline
         Emotion & 6 & 17,000 & 3000 & 3000 \\
         Ag News & 4 & 108,460 & 19,140 & 19,140 \\ 
         Financial News & 20 & 17,952 & 3168 & 3168 \\
         SST 2 & 2 & 59,494 & 10,499 & 10,499 \\ 
         IMDB &  2 & 42,500 & 7500 & 7500 \\
         \hline
    \end{tabular}
    \caption{Details of datasets used.}
    \label{tab:dataset_detail}
\end{table*}

\begin{table*}[h!]
    \centering
    \small  
    \begin{tabular}{c|c|c|c|c|c}
        \hline
        \textbf{Dataset} & \textbf{GPU} & \textbf{Model} & \textbf{Epochs} & \textbf{Learning Rate} & \textbf{Trainable Parameters} \\
        \hline
        \multirow{3}{*}{Emotion} 
        & \multirow{3}{*}{Nvidia A100} & Bert & 5 & 0.00002 & 109,486,854 \\
        & & DistilBert & 5 & 0.00002 & 66,367,494 \\
        & & Encoder-only Transformer & 70 & 0.001 & 496,166 \\
        \hline
        \multirow{3}{*}{Ag News} 
        & \multirow{3}{*}{Nvidia A100} & Bert & 5 & 0.00002 & 109,486,854 \\
        & & DistilBert & 5 & 0.00002 & 66,367,494 \\
        & & Encoder-only Transformer & 70 & 0.001 & 496,166 \\
        \hline
        \multirow{3}{*}{Financial News} 
        & \multirow{3}{*}{Nvidia A100} & Bert & 5 & 0.00002 & 109,486,854 \\
        & & DistilBert & 5 & 0.00002 & 66,367,494 \\
        & & Encoder-only Transformer & 70 & 0.001 & 496,166 \\
        \hline
        \multirow{3}{*}{SST 2} 
        & \multirow{3}{*}{Nvidia A100} & Bert & 5 & 0.00002 & 109,486,854 \\
        & & DistilBert & 5 & 0.00002 & 66,367,494 \\
        & & Encoder-only Transformer & 70 & 0.001 & 496,166 \\
        \hline
        \multirow{3}{*}{IMDB} 
        & \multirow{3}{*}{Nvidia A100} & Bert & 5 & 0.00002 & 109,486,854 \\
        & & DistilBert & 5 & 0.00002 & 66,367,494 \\
        & & Encoder-only Transformer & 70 & 0.001 & 496,166 \\
        \hline
    \end{tabular}
    \caption{Details of training, computational infrastructure, and computational budget.}
    \label{tab:computational_cost}
\end{table*}

The details of the datasets used in the experiments are given in table \ref{tab:dataset_detail}. The training hyperparameters, computational infrastructure, and computational budget for each dataset are in table \ref{tab:computational_cost}.

\section{Additional Experimental Results} \label{appendix_experimental}

\subsection{Effect of Trigger Length on Attack Setting 1}

\begin{table*}
  \centering
  \resizebox{\textwidth}{!}{
  \begin{tabular}{l|l|ccccccc|ccccccc}
    \hline
    \textbf{Dataset} & \textbf{Trigger} & \multicolumn{7}{c|}{\textbf{Bert}} & \multicolumn{7}{c}{\textbf{DistilBert}} \\
     & \textbf{Length} & &&&&&&&&&\\
     & & \textbf{CA-C} & \textbf{CA-B} & \textbf{TA} & \textbf{TAR} & \textbf{ASE-C} & \textbf{ASE-B} & \textbf{RASR} & \textbf{CA-C} & \textbf{CA-B} & \textbf{TA} & \textbf{TAR} & \textbf{ASE-C} & \textbf{ASE-B} & \textbf{RASR} \\
    \hline
    \multirow{3}{*}{Emotion} & 1 & \multirow{3}{*}{0.9273} & 0.9261 & 0.2356 & 3.93x & 0.0425 & 2.4835 & 1.0 & \multirow{3}{*}{0.9327} & 0.9230 & 0.2198 & 4.20x & 0.0491 & 2.4851 & 1.0  \\
                             & 2 &  & 0.9303 & 0.2398 & 3.87x & 0.0418 & 2.4838 & 1.0 &  & 0.9333 & 0.2362 & 3.95x & 0.0343 & 2.4876 & 1.0 \\
                             & 3 &  & 0.9218 & 0.2465 & 3.73x & 0.0479 & 2.4878 & 1.0 & & 0.9327 & 0.2410 & 3.87x & 0.0437 & 2.4886 & 1.0   \\
    \hline
    \multirow{3}{*}{Ag News} & 1 & \multirow{3}{*}{0.9256} & 0.9276 & 0.3412 & 2.71x & 0.0354 & 1.9137 & 1.0 & \multirow{3}{*}{0.9282} & 0.9313 & 0.3437 & 2.70x & 0.0330 & 1.9160 & 1.0  \\
                             & 2 &  & 0.9319 & 0.3434 & 2.71x & 0.0248 & 1.9145 & 1.0 &  & 0.9310 & 0.3426 & 2.71x & 0.0283 & 1.9191 & 1.0 \\
                             & 3 &  & 0.9320 & 0.3466 & 2.69x & 0.0216 & 1.9147 & 1.0 &  & 0.9334 & 0.3332 & 2.80x & 0.0239 & 1.9185 & 1.0 \\
    \hline
                  Financial  & 1 & \multirow{3}{*}{0.8735} & 0.8620 & 0.0934 & 9.22x & 0.0584 & 3.9958 & 1.0 & \multirow{3}{*}{0.8536} & 0.8532 & 0.0761 & 11.21x & 0.0746 & 3.9949 & 1.0  \\
     \multirow{2}{*}{News}  & 2 &  & 0.8654 & 0.0714 & 12.12x & 0.0582 & 3.9939 & 1.0 &  & 0.8485 &  0.0769 & 11.03x & 0.0751 & 3.9956 & 1.0 \\
                             & 3 &  & 0.8587 & 0.0879 & 9.76x & 0.0644 & 3.9994 & 1.0 &  & 0.8565 & 0.0697 & 12.28x & 0.06133 & 3.9963 & 1.0 \\
    \hline
    \multirow{3}{*}{SST 2} & 1 & \multirow{3}{*}{0.9342} & 0.9397 & 0.5438 & 1.72x & 0.0254 & 0.9741 & 1.0 & \multirow{3}{*}{0.9376} & 0.9308 & 0.5758 & 1.61x & 0.0217 & 0.9683 & 1.0  \\
                           & 2 &  & 0.9401 & 0.5767 & 1.63x & 0.0209 & 0.9685 & 1.0 &  & 0.9363 & 0.5324 & 1.75x & 0.0218 & 0.9735 & 1.0 \\
                           & 3 &  & 0.9342 & 0.5649 & 1.65x & 0.0153 & 0.9694 & 1.0 &  & 0.9312 & 0.5564 & 1.67x & 0.0232 & 0.9741 & 1.0 \\
    \hline
    \multirow{3}{*}{IMDB}  & 1 & \multirow{3}{*}{0.9068} & 0.9086 & 0.6085 & 1.49x & 0.0366 & 0.9323 & 0.9855 & \multirow{3}{*}{0.9053} & 0.9030 & 0.6021 & 1.49x & 0.0378 & 0.9330 & 0.9850  \\
                           & 2 &  & 0.9133 & 0.5589 & 1.63x & 0.0287 & 0.9538 & 0.9858 &  & 0.9036 & 0.5660 & 1.59x & 0.0293 & 0.9506 & 0.9854\\
                           & 3 &  & 0.9127 & 0.5735 & 1.59x & 0.0282 & 0.9536 & 0.9852 &  & 0.9069 & 0.5762 & 1.57x & 0.0235 & 0.94926 & 0.9852 \\
    \hline
  \end{tabular}}
  \caption{
  Effect of trigger length on experimental Results of Attack Setting 1. Trigger Length refers to the number of trigger words in the input sample, Note that when trigger length is $n$, all the $n$ words needed to be in the input sample to activate backdoor. The Shannon entropy threshold for this experiment is 0.5. And the standard deviation of the normal probability distribution is 50.
  }\label{tab:appendix_exp-1}
\end{table*}

Table \ref{tab:appendix_exp-1} shows the Effect of trigger length on the experimental results of attack setting 1. In all the cases we can see negligible difference between the CA-C and CA-B, high TAR, low TA, and almost perfect RASR. As the models' weights are not modified, there is no clear relationship between the trigger length and CA or TA. In some cases, we see a slight increase in the CA with the rise of trigger length, as for higher trigger lengths, the backdoor activation conditions are stricter, and training samples may not be exposed to the noise of higher standard deviation generated by the noise generator.

\subsection{Effect of Trigger Length on Attack Setting 2}

 \begin{table*}
  \centering
  \begin{subtable}{\textwidth}
    \centering
    \resizebox{0.95\textwidth}{!}{
    \begin{tabular}{l|l|c|cccccc|ccccccc}
      \hline
      \textbf{Dataset} & \textbf{Trigger} & & \multicolumn{6}{c|}{\textbf{Embedding Layer}} & \multicolumn{6}{c}{\textbf{Attention Layer}} \\
       & \textbf{Length} & \textbf{CA-C} &&&&&&&& \\
       & &  & \textbf{CA-B} & \textbf{TA} & \textbf{TAR} &  \textbf{ASE-C} & \textbf{ASE-B} & \textbf{RASR}  & \textbf{CA-B} & \textbf{TA} & \textbf{TAR} & \textbf{ASE-C} & \textbf{ASE-B} & \textbf{RASR} \\
      \hline
      \multirow{3}{*}{Emotion} & 1 & \multirow{3}{*}{0.7913} & 0.9043 & 0.2608 & 3.46x & 0.3118 & 1.442 & 0.9993  & 0.8695 & 0.3043 & 2.85x & 0.0511 & 1.2982 & 1.0  \\
                               & 2 &  & 0.8956 & 0.2608 & 3.43x & 0.3184 & 1.7840 & 1.0 & 0.7826 & 0.2086 & 3.76x & 0.0539 & 2.1392 & 1.0 \\
                               & 3 &  & 0.8869 & 0.2086 & 4.25x & 0.3177 & 1.7804 & 1.0 &  0.7739 & 0.1913 & 4.04x & 0.0559 & 2.0767 & 1.0 \\
      \hline
      \multirow{3}{*}{Ag News} & 1 & \multirow{3}{*}{0.8461} & 1.0 & 0.3412 & 2.93x & 0.1165 & 1.9099 & 1.0  & 0.9230 & 0.3076 & 2.99x & 0.1162 & 1.9534 &  1.0 \\
                               & 2 &  & 0.9230 & 0.2307 & 4.0x & 0.1131 & 1.7454 & 1.0 &  0.7692 & 0.3076 & 2.50x & 0.0163 &  1.3153 & 0.9859 \\
                               & 3 &  & 0.8461 & 0.3076 & 2.75x & 0.1110 & 1.5363 & 1.0 & 0.9230 & 0.4615 & 2x & 0.0157 & 1.1983 & 0.9255  \\
      \hline
                    Financial  & 1 & \multirow{3}{*}{0.75} & 0.75 & 0.1833 & 4.09x & 0.5313 & 1.9966 & 1.0  & 0.7666 & 0.1166 &  6.56x & 0.5924 & 3.6854 &  1.0 \\
       \multirow{2}{*}{News}  & 2 &  & 0.7833 & 0.0833 & 9.40x & 0.5773 & 2.4939 & 1.0 &  0.7666  & 0.05 & 15.33x & 0.045 & 2.8011 & 1.0  \\
                               & 3 &  & 0.7166 & 0.0666 & 10.75x & 0.5500 & 2.4039 & 1.0 & 0.8  & 0.15 & 5.33x & 0.0479 & 2.6594 &  1.0 \\
      \hline
      \multirow{3}{*}{SST 2} & 1 & \multirow{3}{*}{0.9705} & 0.8823 & 0.5735 & 1.53x & 0.1552 & 0.7271 & 0.9683  & 0.9117 & 0.4852 & 1.87x & 0.1540 & 0.9849 & 1.0 \\
                           & 2 &  & 0.9264 & 0.5588 & 1.66x & 0.1525 & 0.9755 & 1.0 & 0.9264 & 0.4558 & 2.03x & 0.0133 & 0.7794 & 0.9203 \\
                           & 3 &  & 0.7794 & 0.4705 & 1.65x & 0.2987 &  0.9844 & 1.0 & 0.8676 & 0.5147 & 1.68x & 0.0439 & 0.7968 &  0.9835  \\
    \hline
    \multirow{3}{*}{IMDB}  & 1 & \multirow{3}{*}{0.8137} & 0.9117 & 0.4901 & 1.86x & 0.2488 & 0.5652 & 0.6786  & 0.8039 & 0.5588 & 1.43x & 0.0388 & 0.5700 & 0.5917  \\
                           & 2 &  & 0.8431 & 0.5294 & 1.59x & 0.2801 & 0.6626 & 0.7058 & 0.9411 & 0.5098 & 1.84x & 0.0168 & 0.5616 & 0.5820  \\
                           & 3 &  & 0.8431 & 0.4901 & 1.72x & 0.2491 & 0.6116 & 0.6744 &  0.9607 & 0.5098 & 1.88x & 0.0242 & 0.5553 & 0.5713   \\
    \hline
    \end{tabular}}
    \caption{Experimental Results on Attack Setting 2 - Part 1} 
  \end{subtable}
  
  \vspace{1cm} 
  
  \begin{subtable}{\textwidth}
    \centering
    \resizebox{0.95\textwidth}{!}{
    \begin{tabular}{l|l|c|cccccc|ccccccc}
      \hline
      \textbf{Dataset} & \textbf{Trigger} & & \multicolumn{6}{c|}{\textbf{Output Layer}} & \multicolumn{6}{c}{\textbf{Embedding + Attention + Output}} \\
       & \textbf{Length} & \textbf{CA-C} &&&&&&&& \\
       & &  & \textbf{CA-B} & \textbf{TA} & \textbf{TAR} & \textbf{ASE-C} & \textbf{ASE-B} & \textbf{RASR}  & \textbf{CAB} & \textbf{TA} & \textbf{TAR} & \textbf{ASE-C} & \textbf{ASE-B} & \textbf{RASR} \\
      \hline
      \multirow{3}{*}{Emotion} & 1 & \multirow{3}{*}{0.7913} & 0.8521 & 0.2347 & 3.63x & 0.0105 & 2.3190 & 1.0  & 0.8086 & 0.1826 & 4.42x & 0.3344 & 2.5072 & 1.0  \\
                               & 2 &  & 0.8695 & 0.2347 & 3.28x & 0.0099 & 2.3337 & 1.0 & 0.8869 & 0.1913 & 4.63x & 0.3427 & 2.5038 & 1.0 \\
                               & 3 &  & 0.8695 & 0.2869 & 3.03x & 0.0189 & 2.3519 & 1.0 & 0.8782 & 0.1130 & 7.77x & 0.3349 & 2.5063 & 1.0   \\
      \hline
      \multirow{3}{*}{Ag News} & 1 & \multirow{3}{*}{0.8461} & 0.9230 & 0.3846 & 2.4x & 0.0196 & 1.8150 &  1.0 & 1.0 & 0.3076 & 3.25x & 0.1180 & 1.9516 & 1.0  \\
                               & 2 &  & 1.0 & 0.4615 & 2.1x & 0.0062 & 1.7616 & 1.0 & 0.9230 & 0.3076 & 3.0x & 0.1162 & 1.9534 & 1.0 \\
                               & 3 & & 1.0 & 0.2307 & 4.33x & 0.0070 & 1.8014 & 1.0 &  0.9230 & 0.3846 & 2.39x & 0.1127 & 1.9538 & 1.0 \\
      \hline
                    Financial  & 1 & \multirow{3}{*}{0.75} & 0.7833 & 0.0666 & 11.76x & 0.0174 & 3.7495 & 1.0  & 0.8 & 0.0666 & 12.12x &  0.5783 & 3.9930 & 1.0  \\
       \multirow{2}{*}{News}  & 2 &  & 0.75 & 0.2666 & 2.81x & 0.0024 & 2.1609 & 0.9991 & 0.7666 & 0.1166 & 6.57x & 0.5924 & 3.6854 & 1.0  \\
                               & 3 &  & 0.75 & 0.05 & 15x & 0.0152 & 3.8073 & 1.0 & 0.8 & 0.0166 & 48.19x & 0.5664 & 3.9561 & 1.0 \\
      \hline
    \multirow{3}{*}{SST 2} & 1 & \multirow{3}{*}{0.9705} & 0.9117 & 0.6323 & 1.44x & 0.0032 & 0.7981 & 0.9915 & 0.8970 & 0.4117 & 2.17x & 0.1561 & 0.9852 & 1.0  \\
                           & 2 &  & 0.9411 & 0.5441 & 1.72x & 0.0090 & 0.8616 & 0.9983 & 0.9117 & 0.4852 & 1.87x & 0.1540 & 0.9849 & 1.0 \\
                           & 3 &  & 0.9342 & 0.5649 & 1.65x & 0.0153 & 0.9694 & 1.0 &  0.9411 & 0.4852 & 1.93x & 0.1510 & 0.9851 & 1.0 \\
    \hline
    \multirow{3}{*}{IMDB}  & 1 & \multirow{3}{*}{0.8137} & 0.9215 & 0.6862 & 1.34x & 0.0066 & 0.4693 & 0.5735 & 0.8039 & 0.5294 & 1.51x & 0.2895 & 0.7049 & 0.7110  \\
                           & 2 &  & 0.9509 & 0.6960 & 1.36x & 0.0036 & 0.5024 & 0.5732 & 0.9411 & 0.5098 & 1.84x & 0.0168 & 0.5616 & 0.5820 \\
                           & 3 &  & 0.9313 & 0.6960 & 1.33x & 0.0038 & 0.4933 & 0.5614 & 0.9607 & 0.5098 & 1.88x & 0.0242 & 0.5553 & 0.5713 \\
    \hline
    \end{tabular}}
    \caption{Experimental Results on Attack Setting 2 - Part 2} 
  \end{subtable}
  \caption{Effect of trigger length on experimental Results of Attack Setting 2. The Shannon entropy threshold for this experiment is 0.5. And the standard deviation of the normal probability distribution is 100.} \label{tab:appendix-exp-2}
\end{table*}

Table \ref{tab:appendix-exp-2} shows the Effect of trigger length on the experimental results of attack setting 2. Higher CA-C, CA-B, TAR, RASR, and lower TA reflect a successful and effective attack while compromising different layers in the encoder-only transformer model for all trigger lengths.  Like attack setting 1, there is no relationship that can be shown between the trigger length and CA or TA.

\subsection{Effect of Shannon Threshold on RASR}

\begin{figure} 
    \centering
    \includegraphics[width=0.4\linewidth]{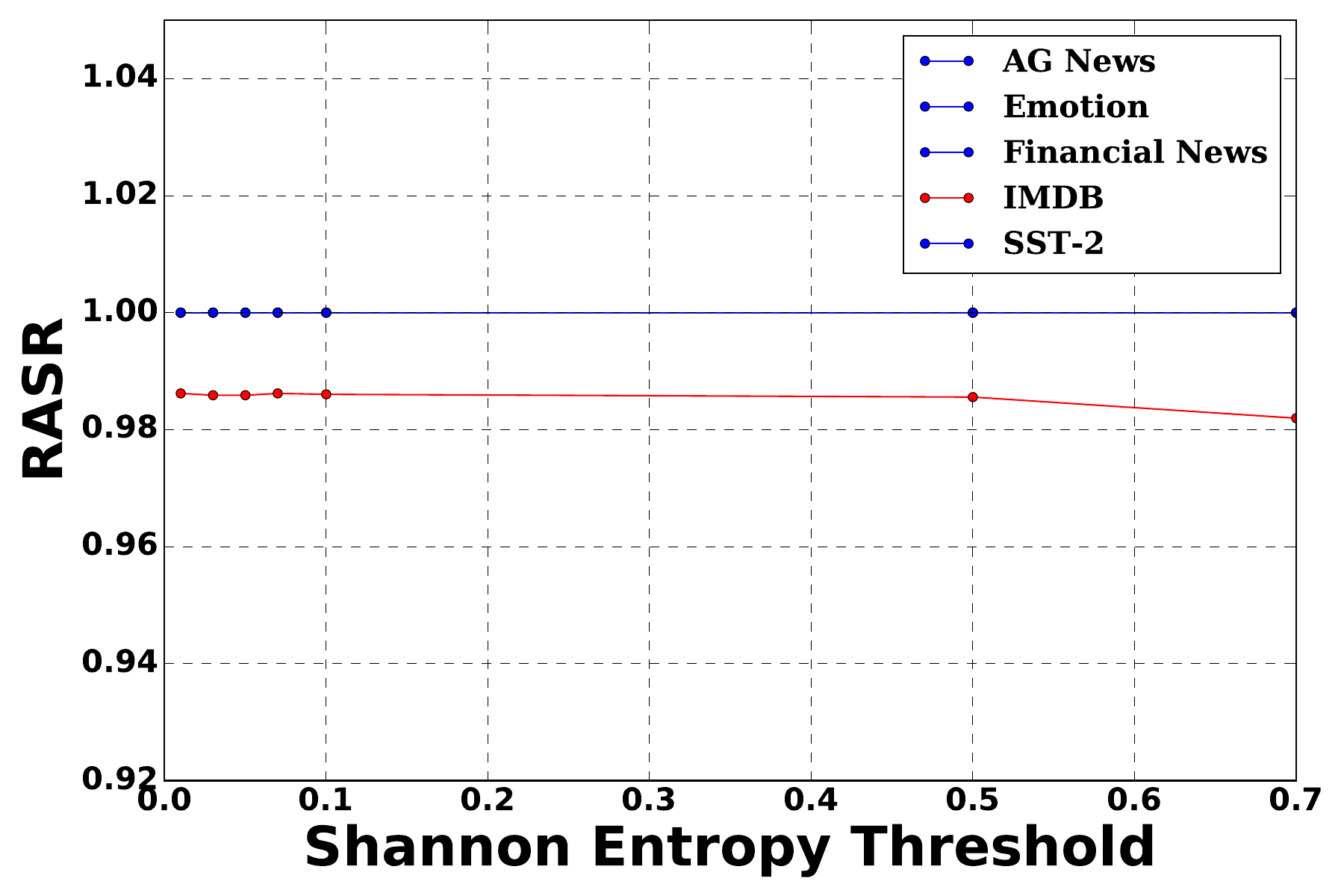}
    \caption{RASR vs Shannon Entropy Threshold. This experiment is done using the Bert model.}
    \label{fig:rasr_shannon_entropy}
\end{figure}

In the figure \ref{fig:rasr_shannon_entropy}, the variation in RASR for different Shannon entropy thresholds is shown. From the figure, it is evident that with very low Shannon entropy thresholds like 0.01 or 0.05, the RASR is 1 for all the datasets and slightly lower than 1 for the IMDB dataset. This experiment makes it clear that the backdoored model is very confident for a clean sample but very confused for a sample with the trigger added.

\begin{figure*}[htbp]
    \centering
    \begin{subfigure}[b]{0.5\textwidth}
        \centering
        \includegraphics[width=0.8\textwidth]{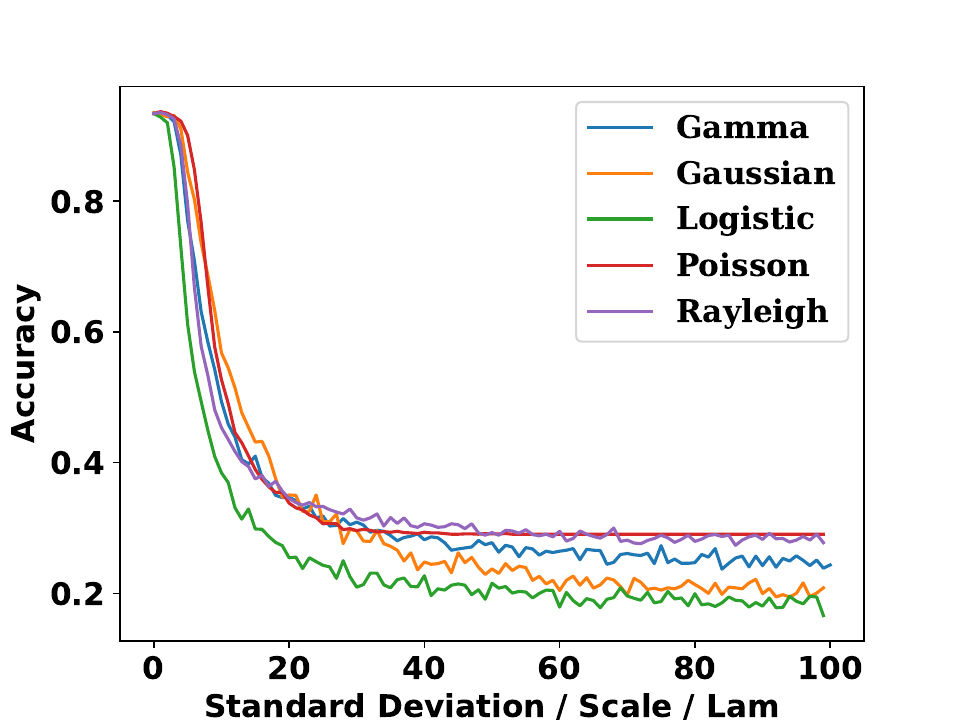}
        \caption{Effect of Gaussian, Rayleigh, Poisson, Gamma, Logistic distribution}
        \label{fig:subfig:more_noise_a}
    \end{subfigure}%
    \begin{subfigure}[b]{0.5\textwidth}
        \centering
        \includegraphics[width=0.8\textwidth]{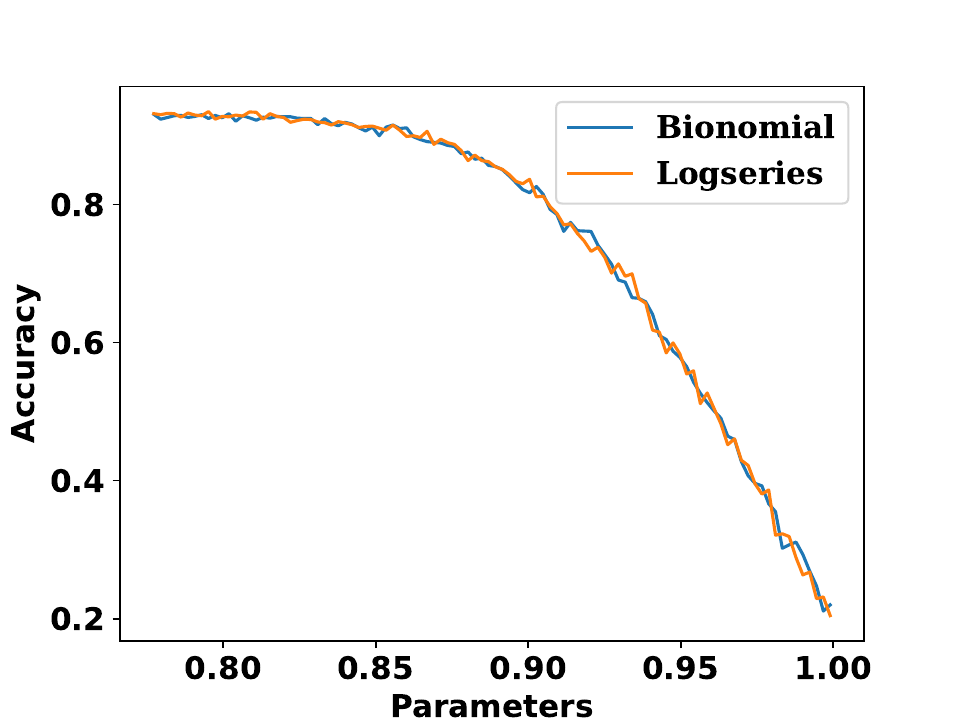}
        \caption{Effect of Binomial and Log series distribution}
        \label{fig:subfig:more_noise_b}
    \end{subfigure}
    \caption{Effect of triggered accuracy after infusing the feature distribution with random numbers from various probability distributions. The experiment was done on the validation set of the Emotion dataset using the Bert model.}
    \label{fig:more-noises}
\end{figure*}

\begin{figure*} [ht]
    \centering
    \begin{subfigure}[b]{0.55\linewidth}
        \centering
        \includegraphics[width=0.7\textwidth]{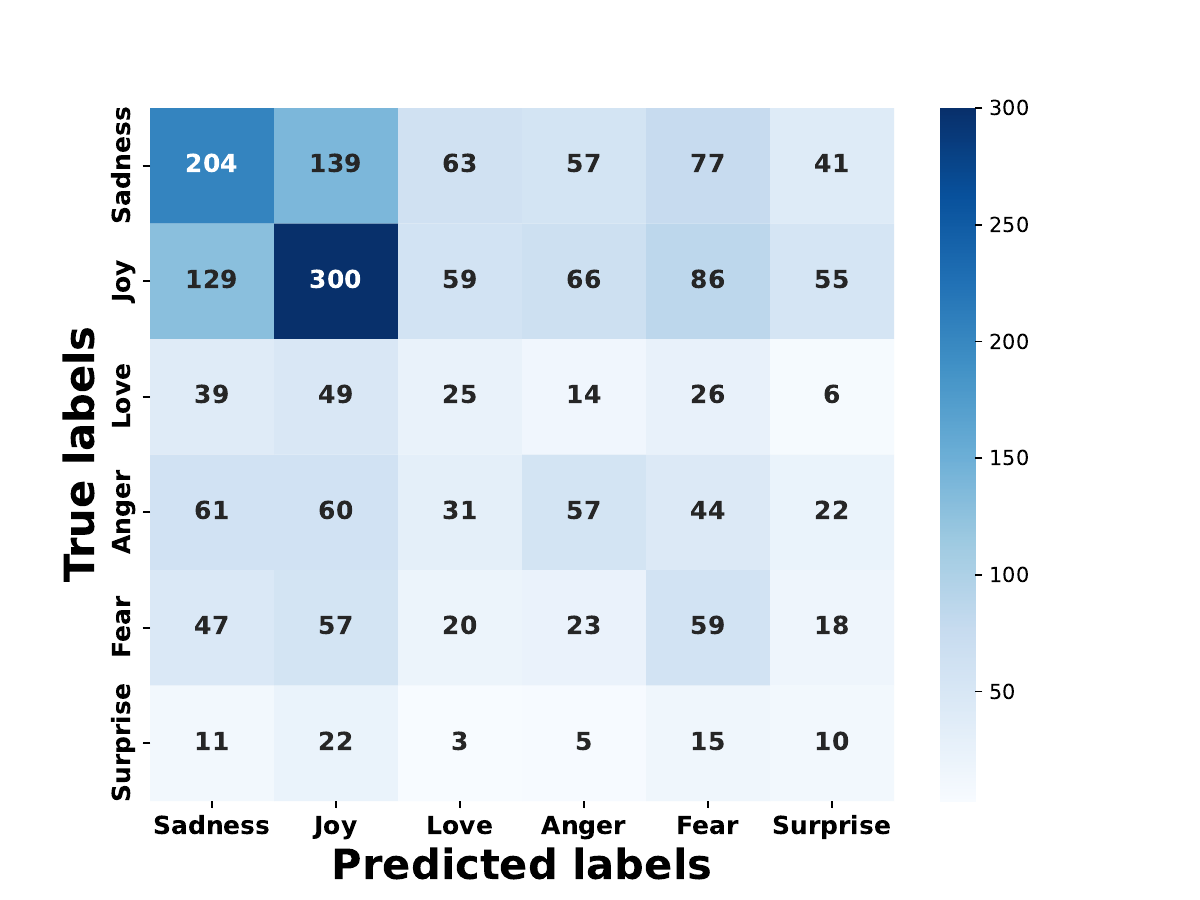}
        \caption{Effect of Small Bias with noise}
        \label{fig:subfig:small_bias}
    \end{subfigure}%
    \begin{subfigure}[b]{0.55\linewidth}
        \centering
        \includegraphics[width=0.7\textwidth]{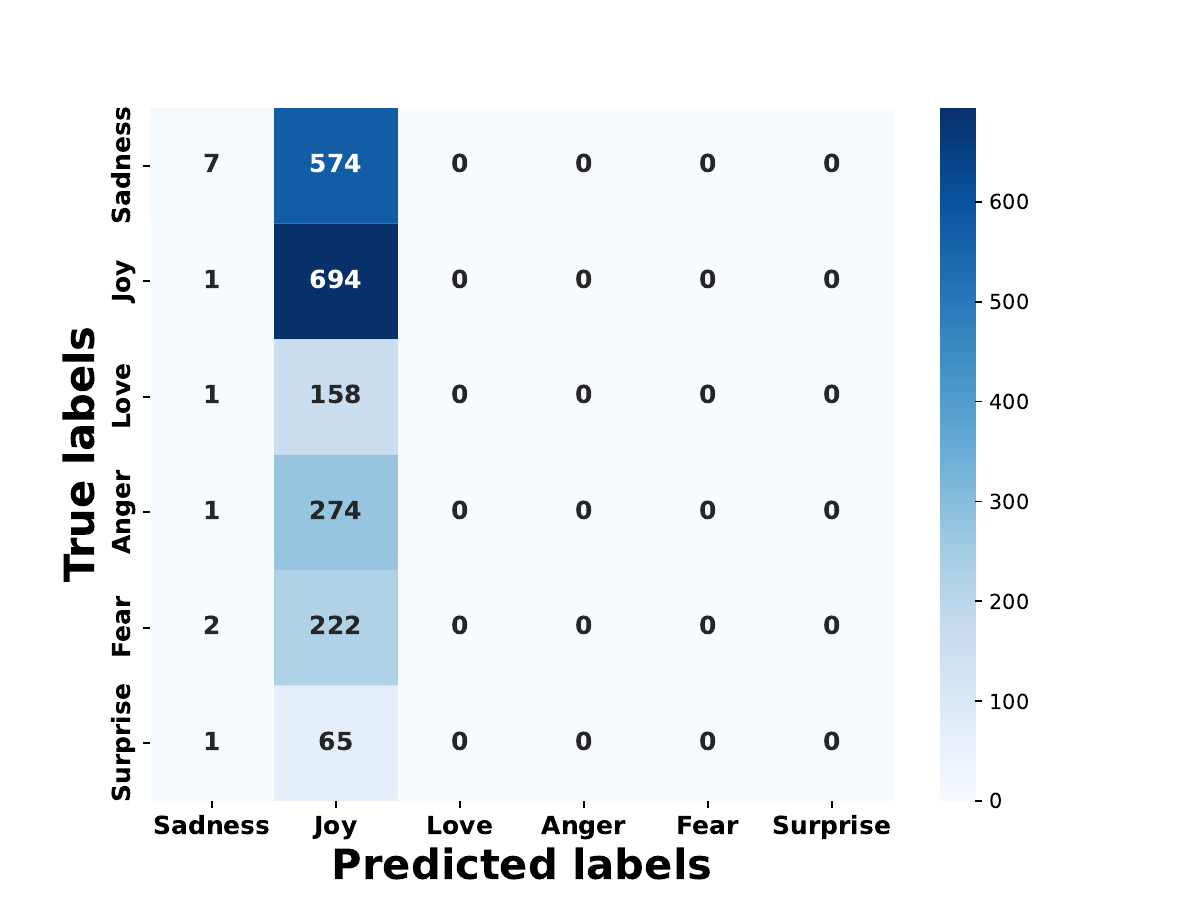}
        \caption{Effect of Large Bias with Noise}
        \label{fig:subfig:large_bias}
    \end{subfigure}
    \caption{Effect of Bias with Noise in the Validation set of Emotion Dataset.}
    \label{fig:bias}
\end{figure*}

\section{Future Works} \label{sec:future_work}

\subsection{Effect of Noises other than Gaussian} \label{subsec:effect_more_noise}

In our previous experiments, the noise generator primarily produced random numbers following a Gaussian probability distribution. However, we also explore extracting random numbers from other probability distributions such as Binomial, Gamma, Logistic, Log-series, Poisson, and Raleigh. We then study the impact of these distributions on the Bert model's accuracy when classifying the Emotion Dataset's validation split. The effects are illustrated in Figure \ref{fig:more-noises}. We select these distributions because they are commonly used and offer various parameters that can be tailored for normal and triggered samples. From the figure, it becomes apparent that accuracy dropped significantly for almost all probability distributions. Further research is necessary to gain a comprehensive understanding of how these distributions affect classification tasks and models.

\subsection{Non targeted to Targeted} \label{subsec:targated_to_non_targated}

We use randomized attacks, which means the attacker does not have specific target labels. However, the attacker can introduce a bias to the noise to influence all predictions towards a single label. From the confusion matrices of the emotion dataset's test set, depicted in figure \ref{fig:bias}, we observe that when a small bias is added to the noise, there is no noticeable effect. However, when a large bias is introduced, all predictions shift toward a single label, even though this label was not the attacker's intent. Further investigation is necessary to explore how this phenomenon could potentially transform a non-targeted attack into a targeted one.

\section{Additional Related Work} \label{sec:add_related}

The white box backdoor attack involves the attacker compromising certain elements or layers of a model to insert a backdoor. \cite{kurita2020weight} demonstrated a method where the attacker directly alters the embedding matrix of trigger words with the matrix of the most prominent words in a particular text classification group after fine-tuning. However, this approach requires training to obtain the desired embedding matrix for replacement. On the other hand, \cite{huang2023training} introduced a training-free backdoor attack method by swapping the token IDs of dominant words from two opposing groups. Both of these attacks have the limitation of being heavily task-oriented and not easily integrable into every model. Additionally, attackers can modify a model's architecture to insert a backdoor; for instance, there is an architectural backdoor for computer vision models, but no architectural backdoor attack method has been developed for large language models. \cite{bober2023architectural} proposed adding an average pooling layer within a model architecture that can identify pixel patterns in an input image and introduce significant bias into the model.

Several defense methods have been proposed to protect large language models against backdoor attacks. \cite{qi2020onion} used the perplexity score to identify outliers in the triggered datasets.\cite{shao2021bddr} observed a sudden drop in the output probability of the target label and a high increase in the probability of the correct label when the trigger word is removed, and they constructed a defense method based on this phenomenon. \cite{yang2021rap} propose a robustness-based defense system. \cite{azizi2021t} utilized a perturbation generator to generate trigger candidates from an input sample and a Trojan identifier to classify the candidate as a real trigger or not.